\documentclass[11pt,superscriptaddress,amsmath,amssymb,nofootinbib,longbibliography,preprint]{revtex4-1}

\usepackage{graphicx}
\usepackage{dcolumn} 
\usepackage{bm}
\usepackage{amssymb}
\usepackage{amsmath} 
\usepackage{color}
\usepackage[normalem]{ulem}

\graphicspath{{./}{figures/}}

\begin{document} 

\title{An explanation for the muon and electron $g-2$ anomalies and dark matter}
\preprint{OU-HET-1054}

\author{Kai-Feng Chen}
\email{B04901029@ntu.edu.tw}
\affiliation{Department of Physics, National Taiwan University, Taipei, Taiwan 10617, R.O.C.}
\affiliation{Institute~of~Astronomy~and~Astrophysics,~Academia~Sinica,~Taipei~10617,~Taiwan}

\author{Cheng-Wei Chiang}
\email{chengwei@phys.ntu.edu.tw}
\affiliation{Department of Physics, National Taiwan University, Taipei, Taiwan 10617, R.O.C.}

\author{Kei Yagyu}
\email{yagyu@het.phys.sci.osaka-u.ac.jp}
\affiliation{Department of Physics, Osaka University, Toyonaka, Osaka 560-0043, Japan}

\begin{abstract}

We propose simple models with a flavor-dependent global $U(1)_\ell$ and 
a discrete $\mathbb{Z}_2$ symmetries to explain the anomalies in the measured anomalous magnetic dipole moments of muon and electron, $(g-2)_{\mu,e}$, while simultaneously accommodating a dark matter candidate. 
These new symmetries are introduced not only to avoid the dangerous lepton flavor-violating decays of charged leptons, but also to ensure the stability of the dark matter. 
Our models can realize the opposite-sign contributions to the muon and electron $g-2$ via one-loop diagrams involving new vector-like leptons. 
Under the vacuum stability and perturbative unitarity bounds as well as the constraints from the dark matter direct searches and related LHC data, we find suitable parameter space to simultaneously explain $(g-2)_{\mu,e}$ and the relic density. 
In this parameter space, the coupling of the Higgs boson with muons can be enhanced by up to $\sim 38\%$ from its Standard Model value, which can be tested in future collider experiments.

\end{abstract}
\maketitle

\section{Introduction}

The Standard Model (SM) for elementary particles has successfully explained a plethora of phenomena in various experiments.  Despite its tremendous success, physics beyond the SM (BSM) is strongly called for to explain neutrino oscillations, dark matter (DM) and baryon asymmetry of the Universe that cannot be accommodated within the SM. 
The question is then how we can experimentally show the existence of such a new physics model.  A discovery of new particles, of course, would provide a direct proof.  However, no report of such discoveries has been given so far, though there is still a possibility for their detection in future collider experiments, such as the High-Luminosity LHC~\cite{Cepeda:2019klc} and the Future Circular Colliders (FCCs)~\cite{Benedikt2019}.  In addition to the direct searches, precision measurements of certain observables can also offer good opportunities to probe new physics (NP). 
Deviations in measured values of the observable from their SM predictions can be attributed to the effects of new particles.

Among various observables, the anomalous magnetic dipole moment of the muon, dubbed the muon $g-2$, has long been thought to be a harbinger for NP~\cite{Czarnecki:2001pv,Giudice:2012ms} and attracted a lot of attention for almost two decades because of the discrepancy between its experimental value measured at Brookhaven National Laboratory (BNL)~\cite{Bennett:2006fi} and the SM expectation.  According to recent studies about the hadronic vacuum polarization contributions~\cite{Keshavarzi:2018mgv,Blum:2018mom,Davier:2019can,Jegerlehner2018} to the muon $g-2$, the discrepancy is at about 3.3$\sigma$ level~\cite{Tanabashi:2018oca}, with the experimental value {\it higher} than the SM prediction.  
See also the recent review on the muon $g-2$, which claims an even larger discrepancy of $3.7\sigma$~\cite{Aoyama:2020ynm}.\footnote{ Note that according to their lattice calculation of the leading order hadronic vacuum polarization contribution to the muon $g-2$, the authors of Ref.~\cite{Borsanyi:2020mff} claim no need of new physics.}
On the other hand, the experimental value of the electron $g-2$ has been updated in 2018~\cite{Parker2018} from a precision determination of the fine-structure constant $\alpha_{\rm em}$.  Interestingly, this measurement also shows a possible disagreement between the data and theory, with the measured value {\it lower} than the SM prediction by about 2.4$\sigma$~\cite{Parker2018}.  These tantalizing opposite deviations have invited many studies to explore suitable NP models~\cite{Liu:2018xkx,Crivellin:2018qmi,Endo:2019bcj,Bauer:2019gfk,Badziak:2019gaf,Abdullah:2019ofw,Hiller:2019mou,Cornella:2019uxs,Haba:2020gkr,Bigaran:2020jil,Jana:2020pxx,Calibbi:2020emz,Yang:2020bmh,Chen:2020jvl,Hati:2020fzp,Dutta:2020scq}.

In order to accommodate both $g-2$ anomalies simultaneously, a characteristic flavor-dependent structure is called for.  In this paper, we propose a new model with a set of new particles whose interactions are constrained by a flavor-dependent global $U(1)_\ell$ symmetry and a $\mathbb{Z}_2$ symmetry, and demonstrate its capabilities to simultaneously accommodate both anomalies and, at the same time, offer a DM candidate.  These new symmetries do not only play an important role in explaining both anomalies, but also forbid dangerous flavor-violating decays of the charged leptons, such as $\mu \to e\gamma$.  Furthermore, they also guarantee the stability of the DM candidate, which is the lightest neutral particle among the new particles.  We find regions in the parameter space that can satisfy the relic density and the direct search constraint of the DM while successfully explaining both $g-2$ anomalies.

This paper is organized as follows. 
In Sec.~\ref{sec:model}, we define our model and give the Yukawa interactions and the scalar potential that are compliant with the symmetries. 
In Sec.~\ref{sec:g-2}, we discuss the new contributions to the muon and electron $g-2$, and scan the parameter space for regions that can explain both anomalies. 
Sec.~\ref{sec:pheno} is devoted to the discussion on DM physics and the collider phenomenology. 
Our conclusion is summarized in Sec.~\ref{sec:conclusion}.

\section{Model\label{sec:model}}

In addition to the SM gauge symmetry $SU(2)_L \otimes U(1)_Y$, our model has an additional global $U(1)_\ell$ and an exact $\mathbb{Z}_2$ symmetries.  The particle content in the lepton and scalar sectors is given in Table~\ref{table:particle}\footnote{Our model can be seen as an extension of the ``SLR'' model proposed in Ref.~\cite{Calibbi:2018rzv}, where only one new fermion is introduced in order to explain the muon $g-2$ anomaly. }.  The lepton sector is comprised of new vector-like isospin singlets $\chi_a$ ($a = e,\mu$) in addition to the SM left- (right-) handed lepton doublets (singlets) $L_L^\ell$ ($\ell_R$) with $\ell=e,\mu,\tau$.  The scalar sector is also extended from the SM one by introducing additional scalar isospin doublet $\eta_D^{}$ and singlet $\eta_S^{}$ fields, with the SM Higgs doublet field denoted by $\Phi$.  All of the new fields ($\chi_a$ and $\eta_{D,S}$) are assigned to be odd under the $\mathbb{Z}_2$ symmetry.  In Table~\ref{table:particle}, the hypercharge $Y_D$ is chosen to be either 0 or 1 in order to include at least one neutral particle in the $\mathbb{Z}_2$-odd sector to be a DM candidate, provided it is the lightest among all the $\mathbb{Z}_2$-odd particles.  For simplicity, we assume $\eta_S^{}$ to be a real field for the scenario with $Y_D = 1$.

\begin{table}[t!h]
\begin{center}
\begin{tabular}{|c||ccc|ccc|}\hline
          & \multicolumn{3}{c|}{Fermion}  &  \multicolumn{3}{c|}{Scalar} \\\hline\hline
 Fields   &  $(L_L^e,L_L^\mu,L_L^\tau)$      & $(e_R,\mu_R,\tau_R)$     & $(\chi_e,\chi_\mu)$    & $\Phi$     & $\eta_D$ & $\eta_S$   \\\hline\hline
$SU(2)_L$ &  ${\bm 2}$   & ${\bm 1}$ & ${\bm 1}$ & ${\bm 2}$ & ${\bm 2}$ & ${\bm 1}$  \\\hline
$U(1)_Y$  & $-1/2$ & $-1$ & $-Y_D$ & $1/2$ & ~~$Y_D-1/2$~~ & $Y_D-1$ \\\hline
$U(1)_\ell$  & ($q_e,q_\mu,1$) & ($q_e,q_\mu,1$) & ($q_e,q_\mu$)  & 0&0&0\\\hline
$\mathbb{Z}_2$  & $+$ &  $+$ & $-$ & $+$ & $-$ & $-$ \\\hline
\end{tabular}
\caption{Particle content and charge assignment under the symmetries $SU(2)_L \otimes U(1)_Y \otimes U(1)_\ell \otimes \mathbb{Z}_2$, where $U(1)_\ell$ is global. The $U(1)_\ell$ charges depend on the lepton flavor with $q_e \neq q_\mu$. 
The parameter $Y_D$ appearing in the hypercharges for $\mathbb{Z}_2$-odd particles can be either 0 or 1. }
\label{table:particle}
\end{center}
\end{table}

The $\mathbb{Z}_2$-even scalar doublet field is parameterized as usual as 
\begin{equation}
\Phi = \begin{pmatrix}
G^+ \\
\frac{1}{\sqrt{2}}(h + v + i G^0)
\end{pmatrix},
\end{equation}
while the $\mathbb{Z}_2$-odd scalar doublet can be parameterized as 
\begin{align}
\begin{split}
\eta_D  &= 
\begin{pmatrix}
\eta^+ \\
\frac{1}{\sqrt{2}}(\eta_H^0 + i \eta_A^0)
\end{pmatrix}~~\text{for}~~Y_D = 1,
\\
\eta_D &= 
\begin{pmatrix}
\frac{1}{\sqrt{2}}(\eta_H^0 + i \eta_A^0) \\
\eta^-
\end{pmatrix}~~\text{for}~~Y_D = 0, 
\end{split}
\end{align}
where $G^\pm$ and $G^0$ are the Nambu-Goldstone bosons that are absorbed as the longitudinal components of the $W^\pm$ and $Z$ bosons, respectively. The vacuum expectation value (VEV) $v$ is fixed by $v = (\sqrt{2}G_F)^{-1/2}$ with $G_F$ being the Fermi decay constant. The VEVs of $\eta_D$ and $\eta_S$ are assumed to be zero in order to avoid spontaneous breakdown of the $\mathbb{Z}_2$ symmetry. The neutral component $h$ in $\Phi$ is identified with the discovered 125-GeV Higgs boson.  Because of the assumed exact $\mathbb{Z}_2$ symmetry, no mixing is allowed between $h$ and the other scalars.  
Hence, the Higgs boson couplings are the same as those of the SM Higgs boson at tree level, while the loop induced couplings such as $h\gamma\gamma$ and $hZ\gamma$ can be modified by loop contributions of the new particles. 
We will discuss the impact of these contributions to the decays of $h \to \gamma\gamma$ and $h \to Z\gamma$ in Sec.~\ref{sec:pheno}. 

The lepton Yukawa interactions and the mass term for $\chi_a$ are given by 
\begin{align} 
{\cal L}_Y 
=& 
\sum_{i=e,\mu,\tau}y_{\rm SM}^i \bar{L}_L^i\ell_R^i \Phi 
+\sum_{a=e,\mu} \left[
f_L^a (\bar{L}_L^a\chi_{R,a})\eta_D   + f_R^a (\bar{\ell}_R^a \chi_{L,a})  \eta_S  + M_{\chi_a} (\bar{\chi}_{L,a}\chi_{R,a})
\right]
+ \text{h.c.} , 
\end{align}
where $(\ell_R^e,\ell_R^\mu,\ell_R^\tau) = (e_R^{},\mu_R^{},\tau_R^{})$.  Because of the $U(1)_\ell$ symmetry, we can naturally realize the flavor-diagonal couplings $f_L$ and $f_R$, so that contributions from the new particles to lepton flavor-violating processes such as $\mu \to e\gamma$ can be avoided at all orders.  It should be emphasized here that analogous to the GIM mechanism, this structure cannot be achieved in a model with only one vector-like lepton, where it is impossible to accommodate both muon and electron $g-2$ while suppressing the $\mu \to e\gamma$ decay to the level consistent with the current experimental bound.  In general, the new Yukawa couplings $f_{L,R}^a$ can be complex, but we assume them to be real for simplicity in the following discussions.  The Lagrangian for the quark and gauge sectors are the same as in the SM.

The most general form of the scalar potential consistent with all the symmetries is given by 
\begin{align} 
V 
=& 
-\mu_\Phi^2|\Phi|^2 + \mu_D^2|\eta_D|^2 +  \mu_S^2|\eta_S|^2\notag\\
& 
+ \frac{\lambda_1}{2} |\Phi|^4 + \frac{\lambda_2}{2} |\eta_D|^4 
  + \lambda_3 |\Phi|^2|\eta_D|^2 + \lambda_4 |\Phi^\dagger \eta_D|^2  + \left[\frac{\lambda_5}{2} \left(\Phi \cdot \eta_D \right)^2 + \text{h.c.} \right]\notag\\
& 
+ \frac{\lambda_6}{2}|\eta_S|^4  + \lambda_7 |\Phi|^2|\eta_S|^2  + \lambda_8 |\eta_D|^2|\eta_S|^2 + [\kappa (\eta_D^\dagger\Phi \eta_S) + \text{h.c.}], 
\end{align}
where 
\begin{align} 
\Phi \cdot \eta_D =
\begin{cases} 
\Phi^\dagger \eta_D & {\rm for}~~Y_D = 1 ~, 
\\
\Phi^T (i\tau_2) \eta_D & {\rm for}~~Y_D = 0 ~,
\end{cases}
\end{align}
with $\tau_2$ being the second Pauli matrix. The phases of $\lambda_5$ and $\kappa$ parameters can be removed by a redefinition of the scalar fields without loss of generality. Therefore, CP symmetry is preserved in the scalar potential. We require $\mu^2_{\Phi}, \mu^2_{D}, \mu^2_{S} > 0$ in order to preserve the stability of the SM vacuum.

The squared mass of the Higgs boson $h$ is given by $m_h^2  = v^2\lambda_1$ in both scenarios of $Y_D = 1$ and $Y_D = 0$.  On the other hand, the mass formulas for the $\mathbb{Z}_2$-odd scalar bosons are different in the two scenarios.  For the scenario with $Y_D = 1$, the singlet field $\eta_S$ is neutral ($\eta_S^0 \equiv \eta_S^{}$), so that the $\eta_H^{0}$ and $\eta_S^0$ fields can mix with each other.  By introducing a mixing angle $\theta$, the mass eigenstates of these neutral scalar fields can be defined through 
\begin{align}
\begin{pmatrix}
\eta_H^0\\
\eta_S^0
\end{pmatrix}
= 
\begin{pmatrix}
c_\theta & -s_\theta \\
s_\theta &  c_\theta 
\end{pmatrix}
\begin{pmatrix}
\eta_1^0\\
\eta_2^0
\end{pmatrix}, 
\end{align}
where $s_\theta \equiv \sin\theta$ and $c_\theta \equiv \cos\theta$.  The mixing angle can be expressed as 
\begin{align}
\tan2\theta &= \frac{2({\cal M}_{H}^2)_{12}}{({\cal M}_H^2)_{11}-({\cal M}_{H}^2)_{22}},  \label{eq:mass3}
\end{align}
where ${\cal M}^2_H$ is the mass matrix in the basis of $(\eta_H^0,\eta_S^0)$: 
\begin{align}
{\cal M}_H^2 = \begin{pmatrix}
\mu_D^2 + \frac{v^2}{2}(\lambda_3 +  \lambda_4 + \lambda_5) & v\kappa \\
v\kappa & 2\mu_S^2 + v^2 \lambda_7
\end{pmatrix}. 
\end{align}
The squared masses of the scalar bosons are then given by 
\begin{align}
\begin{split}
m_{\eta^\pm}^2 & = \mu_D^2 + \frac{v^2}{2}\lambda_3 , \\
m_{\eta_A}^2  & = \mu_D^2 + \frac{v^2}{2}(\lambda_3 + \lambda_4 - \lambda_5), \\
m_{\eta_1}^2 &= c_\theta^2 ({\cal M}_{H}^2)_{11} + s_\theta^2 ({\cal M}_H^2)_{22} + s_{2\theta} ({\cal M}_{H}^2)_{12},  
\\
m_{\eta_2}^2 &= s_\theta^2 ({\cal M}_{H}^2)_{11} + c_\theta^2 ({\cal M}_H^2)_{22} - s_{2\theta} ({\cal M}_{H}^2)_{12}.  
\end{split}
\end{align}
From the above expressions, we can write the parameters in the scalar potential in terms of the physical parameters as follows: 
\begin{align}
\label{eq:potential_param_YD=0}
\begin{split}
\mu_D^2 &= m_{\eta^\pm}^2 -\frac{v^2}{2}\lambda_3, \\
\mu_S^2 &= \frac{1}{2}(m_{\eta_1}^2s_\theta^2 + m_{\eta_2}^2c_\theta^2 - v^2\lambda_7), \\
\lambda_4 &= \frac{1}{v^2}(m_{\eta_1}^2c_\theta^2 + m_{\eta_2}^2s_\theta^2 +m_{\eta_A}^2 -2m_{\eta^\pm}^2), \\
\lambda_5 &= \frac{1}{v^2}(m_{\eta_1}^2c_\theta^2 + m_{\eta_2}^2s_\theta^2 - m_{\eta_A}^2), \\
\kappa & = \frac{1}{v}s_\theta c_\theta (m_{\eta_1}^2 - m_{\eta_2}^2). 
\end{split}
\end{align}
After fixing $m_h$ and $v$ to their experimental values, the remaining ten independent parameters in the scalar potential are then chosen to be
\begin{align}
m_{\eta^\pm},\quad 
m_{\eta_A},\quad
m_{\eta_1},\quad
m_{\eta_2},\quad
\theta,\quad \lambda_3,\quad \lambda_7, 
\label{eq:free-para-YD1}
\end{align}
and the quartic couplings $\lambda_{2,6,8}$ for the $\mathbb{Z}_2$-odd scalar bosons.

For the scenario with $Y_D = 0$, the singlet field $\eta_S^{}$ is singly-charged ($\eta_S^\pm \equiv \eta_S^{}$), so that the charged components of the inert doublet field $\eta^\pm$ can mix with $\eta_S^\pm$.  Similar to the above scenario, the mass eigenstates are defined through 
\begin{align}
\begin{pmatrix}
\eta^\pm\\
\eta_{S}^\pm
\end{pmatrix}
= 
\begin{pmatrix}
c_\theta & -s_\theta \\
s_\theta &  c_\theta 
\end{pmatrix}
\begin{pmatrix}
\eta_1^\pm\\
\eta_2^\pm
\end{pmatrix}, 
\end{align}
with 
\begin{align}
\tan2\theta &= \frac{2({\cal M}_\pm^2)_{12}}{({\cal M}_\pm^2)_{11}-({\cal M}_\pm^2)_{22}}. 
\end{align}
The mass matrix ${\cal M}_\pm^2$ is expressed in the basis of $(\eta^{\pm},\eta_S^{\pm})$ as 
\begin{align}
{\cal M}_\pm^2 = \begin{pmatrix}
\mu_D^2 + \frac{v^2}{2}(\lambda_3 +  \lambda_4 ) & \frac{v\kappa}{\sqrt{2}} \\
\frac{v\kappa}{\sqrt{2}}& \mu_S^2 + \frac{v^2}{2} \lambda_7
\end{pmatrix}. 
\end{align}
The squared masses of the scalar fields are then given by 
\begin{align}
\begin{split}
m_{\eta_1^\pm}^2 &= c_\theta^2 ({\cal M}_\pm^2)_{11} + s_\theta^2 ({\cal M}_\pm^2)_{22} + s_{2\theta} ({\cal M}_\pm^2)_{12}  , \\
m_{\eta_2^\pm}^2 &= s_\theta^2 ({\cal M}_\pm^2)_{11} + c_\theta^2 ({\cal M}_\pm^2)_{22} - s_{2\theta} ({\cal M}_\pm^2)_{12} , \\
m_{\eta_A}^2  & = \mu_D^2 + \frac{v^2}{2}(\lambda_3  - \lambda_5), \\
m_{\eta_H}^2  & = \mu_D^2 + \frac{v^2}{2}(\lambda_3  + \lambda_5). 
\end{split}
\end{align}
Some of the parameters in the potential can be rewritten in terms of the physical parameters as 
\begin{align}
\begin{split}
\mu_D^2 & = \frac{1}{2}(m_{\eta_A}^2 +  m_{\eta_H}^2 -v^2\lambda_3), \\ 
\mu_S^2 & = m_{\eta_1^\pm}^2c_\theta^2 + m_{\eta_2^\pm}^2s_\theta^2 - \frac{v^2}{2}\lambda_7, \\
\kappa & = \frac{\sqrt{2}}{v}s_\theta c_\theta(m_{\eta_1^\pm}^2 - m_{\eta_2^\pm}^2), \\
\lambda_4 & = -\frac{1}{v^2}(m_{\eta_A}^2 + m_{\eta_H}^2 -2m_{\eta_1^\pm}^2 c_\theta^2 -2m_{\eta_2^\pm}^2 s_\theta^2), \\
\lambda_5 & = \frac{1}{v^2}(m_{\eta_H}^2 - m_{\eta_A}^2). 
\end{split}
\end{align}
Therefore, the ten independent parameters in the scalar potential can be chosen as 
\begin{align}
m_{\eta_1^\pm},\quad 
m_{\eta_2^\pm},\quad 
m_{\eta_A},\quad
m_{\eta_H},\quad
\theta,\quad
\lambda_{3,7},
\label{eq:free-para-YD0}
\end{align}
and the quartic couplings $\lambda_{2,6,8}$ for the inert scalar fields.

The parameters in the scalar potential are subject to the constraints of perturbative unitarity and vacuum stability. In order for our models to be perturbative, we require all the quartic couplings $\lambda_i$ in the potential to satisfy
\begin{equation}
    \label{eq:per_bound}
    \frac{\lambda^2_i}{4\pi} < 1.
\end{equation}
To impose the tree-level unitarity constraints, we consider all possible $2\to 2$ elastic scatterings for the bosonic states in the high energy limit, and obtain thirteen independent eigenvalues of the $s$-wave amplitude matrix, expressed in terms of the scalar quartic couplings. By demanding the magnitude of each eigenvalue to be smaller than 8$\pi$~\cite{Gunion:1989we}, we find the following conditions for the quartic couplings\footnote{The quartic terms of the scalar potential have the same forms as those given in the so-called next-to-two-Higgs doublet model studied in Ref.~\cite{Muhlleitner:2016mzt} except for notational differences.  We have confirmed that our results are consistent with those given in Ref.~\cite{Muhlleitner:2016mzt}.}; 
\begin{align}
&\left|\frac{1}{2}\left(\lambda_{1}+\lambda_{2}+\sqrt{\left(\lambda_{1}-\lambda_{2}\right)^{2}+4 \lambda_{4}^{2}}\right) \right| <8\pi, \label{eq:per-lam2_1}\\
&\left|\frac{1}{2}\left(\lambda_{1}+\lambda_{2}+\sqrt{\left(\lambda_{1}-\lambda_{2}\right)^{2}+4 \lambda_{5}^{2}}\right)\right|  <8\pi, \label{eq:per-lam2_2}\\
&\left|\lambda_3 + 2 \lambda_4 \pm 3 \lambda_5\right|<8 \pi, \quad
\left|\lambda_3 \pm \lambda_5\right| <8 \pi, \quad
\left|\lambda_3 \pm \lambda_4\right| <8 \pi, \quad c_1|\lambda_{7,8}| <8 \pi, \label{eq:per-lam3_1}\\
&|a_{1,2,3}| <8 \pi, \label{eq:per-lam268}
\end{align}
where $a_{1,2,3}$ are the eigenvalues for the following $3\times 3$ matrix
\begin{align}
\begin{pmatrix}
3\lambda_1 & 2\lambda_3 + \lambda_4 & c_2\lambda_7 \\
2\lambda_3 + \lambda_4 & 3\lambda_2 & c_2\lambda_8\\
c_2\lambda_7  & c_2\lambda_8  & c_3\lambda_6
\end{pmatrix}, 
\end{align}
with the coefficients $(c_1,c_2,c_3) = (2,2,6)$ for $Y_D = 1$ and $(c_1,c_2,c_3) = (1,\sqrt{2},2)$ for $Y_D = 0$. 
If we take $\lambda_{6,7,8} = 0$, the above expressions are reduced to those in the two-Higgs doublet model (see, e.g., Ref.~\cite{Kanemura:2004mg}). 

To ensure the stability of the SM vacuum, besides requiring the quadratic terms $\mu^2_{D}$ and $\mu^2_{S}$ to be positive, we further require the potential to be bounded from below. The bounded-from-below conditions are given by~\cite{Muhlleitner:2016mzt} 
\begin{equation}
\lambda_i \in \Omega_1\cup\Omega_2,\quad i = 1, \ldots, 8
\end{equation}
where 
\begin{align}
&
\Omega_{1}=\Big\{\lambda_{1}, \lambda_{2}, \lambda_{6}>0 ; \sqrt{\lambda_{1} \lambda_{6}}+\lambda_{7}>0 ; \sqrt{\lambda_{2} \lambda_{6}}+\lambda_{8}>0 ; \notag \\
& \qquad\qquad
\sqrt{\lambda_{1} \lambda_{2}}+\lambda_{3}+D>0 ; \lambda_{7}+\sqrt{\frac{\lambda_{1}}{\lambda_{2}}} \lambda_{8} \geq 0\Big\},
\label{eq:stability1}
\\
&
\Omega_{2}=\Big\{\lambda_{1}, \lambda_{2}, \lambda_{6}>0 ; \sqrt{\lambda_{2} \lambda_{6}} \geq \lambda_{8}>-\sqrt{\lambda_{2} \lambda_{6}} ; \sqrt{\lambda_{1} \lambda_{6}}>-\lambda_{7} \geq \sqrt{\frac{\lambda_{1}}{\lambda_{2}}} \lambda_{8} ; \notag \\
& \qquad\qquad
\sqrt{\left(\lambda_{7}^{2}-\lambda_{1} \lambda_{6}\right)\left(\lambda_{8}^{2}-\lambda_{2} \lambda_{6}\right)}>\lambda_{7} \lambda_{8}-\left(D+\lambda_{3}\right) \lambda_{6}\Big\}
\label{eq:stability2}
\end{align}
in which $D = \max\left\{0, \lambda_4 - \lambda_5 \right\}$.

For the convenience of discussions, we define the scalar trilinear coupling $\lambda_{\phi_1\phi_2\phi_3}$ to be the coefficient of the $\phi_1\phi_2\phi_3$ term in the Lagrangian, where $\phi_i$ are the physical scalar bosons in our model.

Before closing this section, we briefly comment on neutrino masses in our model.  Under the charge assignments given in Table~\ref{table:particle}, the structure of the dimension-5 operator is strongly constrained: only $\overline{L_L^{\tau c}}\Phi (\Phi^c)^\dagger L_L^\tau$ is allowed.  In order to obtain nonzero values for all the elements of the $3\times 3$ neutrino mass matrix for the observed mixing pattern, two additional Higgs doublet fields, denoted by $\Phi_e$ and $\Phi_\mu$, are required.  Taking the $U(1)_\ell$ charge for $\Phi_e$ and $\Phi_\mu$ to be $-q_e$ and $-q_\mu$, respectively, we can write down all the dimension-5 effective Lagrangian as 
\begin{align} 
{\cal L}_{\rm eff} = \sum_{i,j=e,\mu,\tau}\frac{c_{ij}}{\Lambda}\overline{L_L^{ic}}\Phi_i (\Phi_i^c)^\dagger L_L^j, 
\end{align} 
where $\Phi_\tau = \Phi$, and $c_{ij}$ and $\Lambda$ are respectively dimensionless couplings and the cutoff scale.  Note that if we consider the case with one of the three Higgs doublets being absent, the neutrino mass matrix has the texture with three zeros; that is, one diagonal and two off-diagonal elements including their transposed elements are zero.  It has been known that such textures cannot accommodate the current neutrino oscillation data~\cite{Xing:2004ik}.  Hence, at least three Higgs doublets are required.  In the following discussions, we consider the model defined with just the Higgs doublet in Table~\ref{table:particle} by assuming the $\Phi_e$ and $\Phi_\mu$ fields to be completely decoupled.

\section{Muon/Electron magnetic dipole moments \label{sec:g-2}}

The anomalous magnetic dipole moment of lepton $\ell$ is usually denoted by $a_\ell \equiv (g-2)_\ell/2$.
Currently, the differences between the experimental value $a_\ell^{\rm exp}$ and the SM prediction $a_\ell^{\rm SM}$ for $\ell = \mu, e$ are given by 
\begin{align}
&\Delta a_\mu \equiv a_\mu^{\rm exp} - a_\mu^{\rm SM} = 261(79)\times 10^{-11}, 
\\
&\Delta a_e \equiv a_e^{\rm exp} - a_e^{\rm SM} = -88(36)\times 10^{-14}, 
\end{align}
presenting about 3.3$\sigma$~\cite{Tanabashi:2018oca} and 2.4$\sigma$~\cite{Parker2018} deviations, respectively. 

\begin{figure}[t]
\begin{center}
 \includegraphics[width=75mm]{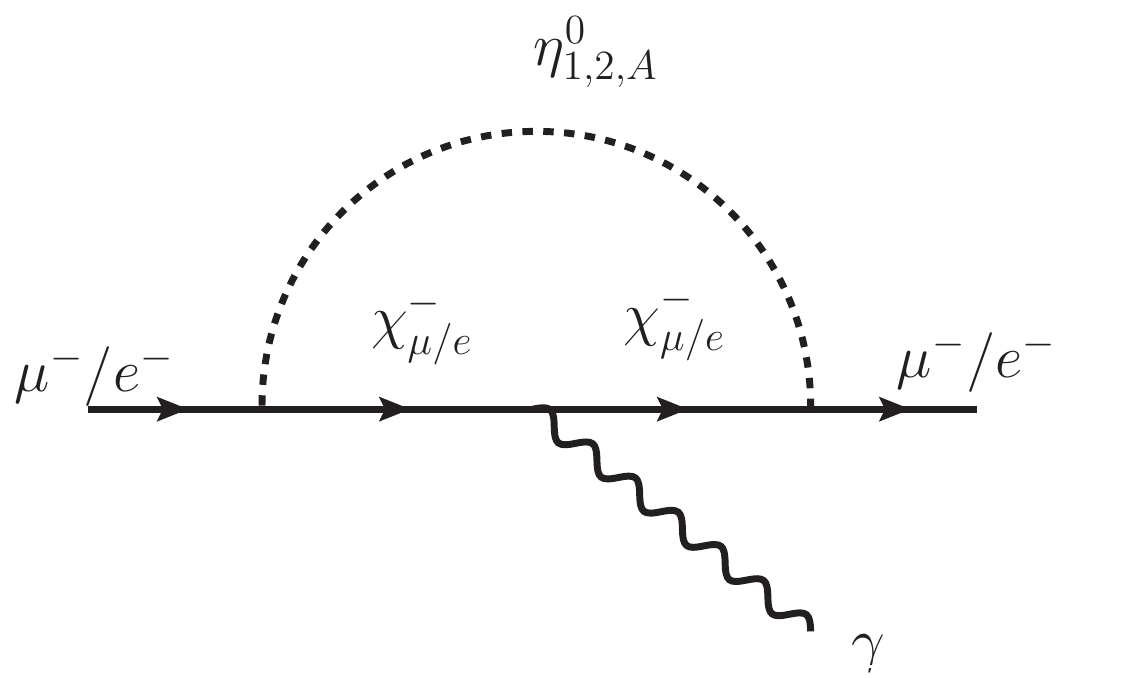} \hspace{5mm}
 \includegraphics[width=75mm]{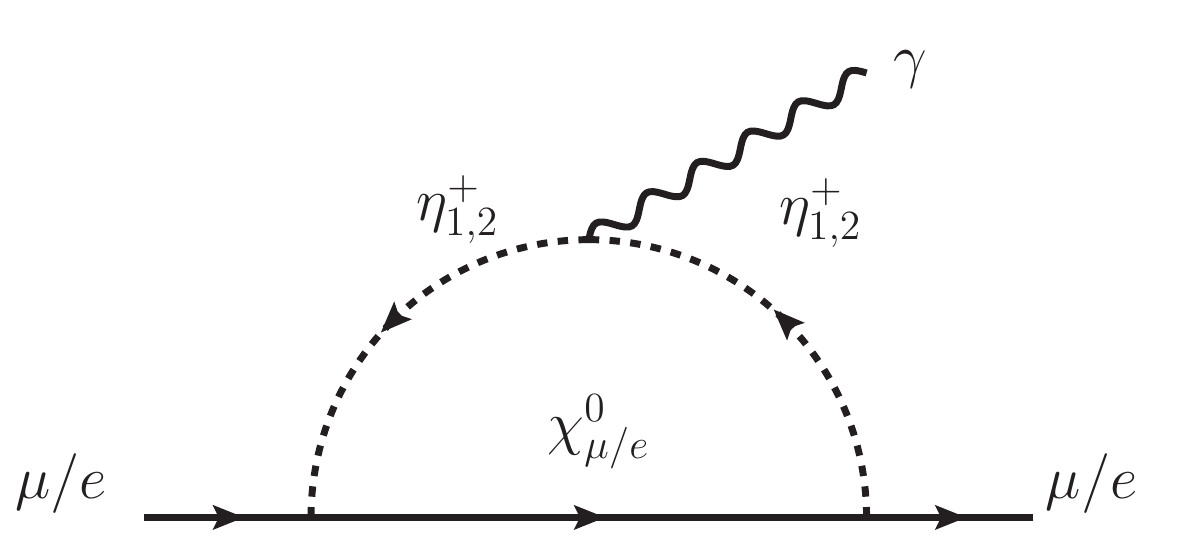}
   \caption{Feynman diagrams for the muon/electron $g-2$.
The left (right) diagram contributes to $g-2$ in the model with $Y_D = 1$ ($Y_D=0$).}
   \label{fig:gminus2}
\end{center}
\end{figure}

In our model, the new contribution to $a_\ell$, denoted by $\Delta a_\ell^{\rm NP}$, mainly comes from the one-loop diagrams shown in Fig.~\ref{fig:gminus2}, with $\mathbb{Z}_2$-odd particles running in the loop. 
These contributions are calculated to be 
\begin{align}
\Delta a_\ell^{\rm NP} &=  -\frac{1}{16\pi^2} \sum_{k = 1,2} \Bigg[\frac{m_\ell^2}{M_{\chi_\ell}^2} (|g_L^{\ell,k}|^2  +|g_R^{\ell,k}|^2)F_2\left(\frac{m_{\eta_k}^2}{M_{\chi_\ell}^2}\right)  \notag\\
&\quad +\frac{2m_\ell}{M_{\chi_\ell}}\text{Re}(g_L^{\ell,k}g_R^{\ell,k*})F_1\left(\frac{m_{\eta_k}^2}{M_{\chi_\ell}^2}\right) \Bigg] 
{
- \frac{|f_L^\ell|^2}{32\pi^2}\frac{m_\ell^2}{M_{\chi_\ell}^2}F_2\left(\frac{m_{\eta_A}^2}{M_{\chi_\ell}^2}\right)} 
\quad  (\text{for}~~Y_D = 1), \label{eq:da1} \\
\Delta a_\ell^{\rm NP} &=-\frac{1}{16\pi^2}\sum_{k = 1,2}\Bigg[\frac{m_\ell^2}{m_{\eta_k^\pm}^2}(|g_L^{\ell,k}|^2  +|g_R^{\ell,k}|^2)F_2\left( \frac{M_{\chi_\ell}^2}{m_{\eta_k^\pm}^2}\right) \notag\\
&\quad\quad\quad\quad\quad\quad\quad +\frac{2M_{\chi_\ell} m_\ell}{m_{\eta_k^\pm}^2}\text{Re}(g_L^{\ell,k} g_R^{\ell,k*})F_3\left(\frac{M_{\chi_\ell}^2}{m_{\eta_k^\pm}^2}\right) \Bigg]\quad  (\text{for}~~Y_D = 0), \label{eq:da2}
\end{align}
where $g_{L,R}^{\ell,k}$ denote the Yukawa couplings for the $\bar{\chi}_\ell \, P_{L,R}\,  \ell \eta_k$ ($\bar{\chi}_\ell \, P_{L,R}\,  \ell \eta_k^\pm$) vertices in the model with $Y_D =1$ (0). 
More explicitly,
\begin{align}
&g_{L}^{\ell,1} = \frac{f_{L}^\ell}{\sqrt{2}} c_\theta,\quad  
g_{L}^{\ell,2} = -\frac{f_{L}^\ell}{\sqrt{2}} s_\theta,\quad
g_{R}^{\ell,1} =  f_R^\ell s_\theta,\quad  g_{R}^{\ell,2}  =  f_R^\ell c_\theta~~
{(\text{for}~~Y_D = 1),} \notag\\
&
{
g_{L}^{\ell,1} = f_{L}^\ell c_\theta,\quad  
g_{L}^{\ell,2} = -f_{L}^\ell s_\theta,\quad
g_{R}^{\ell,1} =  f_R^\ell s_\theta,\quad  g_{R}^{\ell,2}  =  f_R^\ell c_\theta~~(\text{for}~~Y_D = 0).}
\label{g-coupling}
\end{align}
The loop functions are defined as follows: 
\begin{align}
\begin{split}
F_1(x) & =\frac{1-4x+3x^2-2x^2\ln x}{2(1-x)^3},\\
F_2(x) & =\frac{1-6x+3x^2+2x^3-6x^2\ln x}{6(1-x)^4},\\
F_3(x) & =\frac{1-x^2+2x\ln x}{2(1-x)^3},
\end{split}
\end{align}
where at any given $x$, we have $F_1(x) \geq F_3(x) > F_2(x)$.  In both Eqs.~\eqref{eq:da1} and \eqref{eq:da2}, the coefficient of $\text{Re}(g_L^{\ell,k}g_R^{\ell,k*})$ can be much larger than that of $|g_{L}^{\ell,k}|^2 + |g_{R}^{\ell,k}|^2$ by a factor of $M_{\chi_\ell}/m_{\ell}$, and becomes the dominant factor for $\Delta a_\ell^{\rm NP}$. 
We note that for a fixed value of $M_{\chi_\ell}$ and the Yukawa couplings, a larger magnitude of the dominant term is obtained for a smaller mass of the scalar boson running in the loop. 
In addition, the contribution to the dominant term from the lighter scalar boson ($\eta_1^0$ or $\eta_1^\pm$) is opposite in sign to that from the heavier one ($\eta_2^0$ or $\eta_2^\pm$) due to the orthogonal rotation of the scalar fields, as seen in Eq.~(\ref{g-coupling}). 
Therefore, the sign of $\Delta a_\ell^{\rm NP}$ is determined by $\text{Re}(g_L^{\ell,1}g_R^{\ell,1*})$.
We thus take $\text{Re}(g_L^{\mu,1}g_R^{\mu,1*}) < 0$ and $\text{Re}(g_L^{e,1}g_R^{e,1*}) > 0$ in order to obtain $\Delta a_\mu^{\rm NP} >0$ and $\Delta a_e^{\rm NP} < 0$, as required by data. 
This in turn can be realized by taking $f^{\mu}_L > 0$, $f^{\mu}_R < 0$, $f^e_{L, R}>0$, and the mixing angle $\theta$ to be in the first quadrant. 
Note here that with a degenerate mass for $\eta_1$ and $\eta_2$, $\Delta a_\ell^{\rm NP}$ would vanish due to the cancellation between the contributions of the two scalar bosons.  
Therefore, a non-zero mass splitting between $\eta_1$ and $\eta_2$ is required. 
For simplicity, we take $|f_L^\ell| = |f_R^\ell| (\equiv f^\ell)$ in the following analyses. 

\begin{figure}
    \centering
    \includegraphics[width=.45\textwidth]{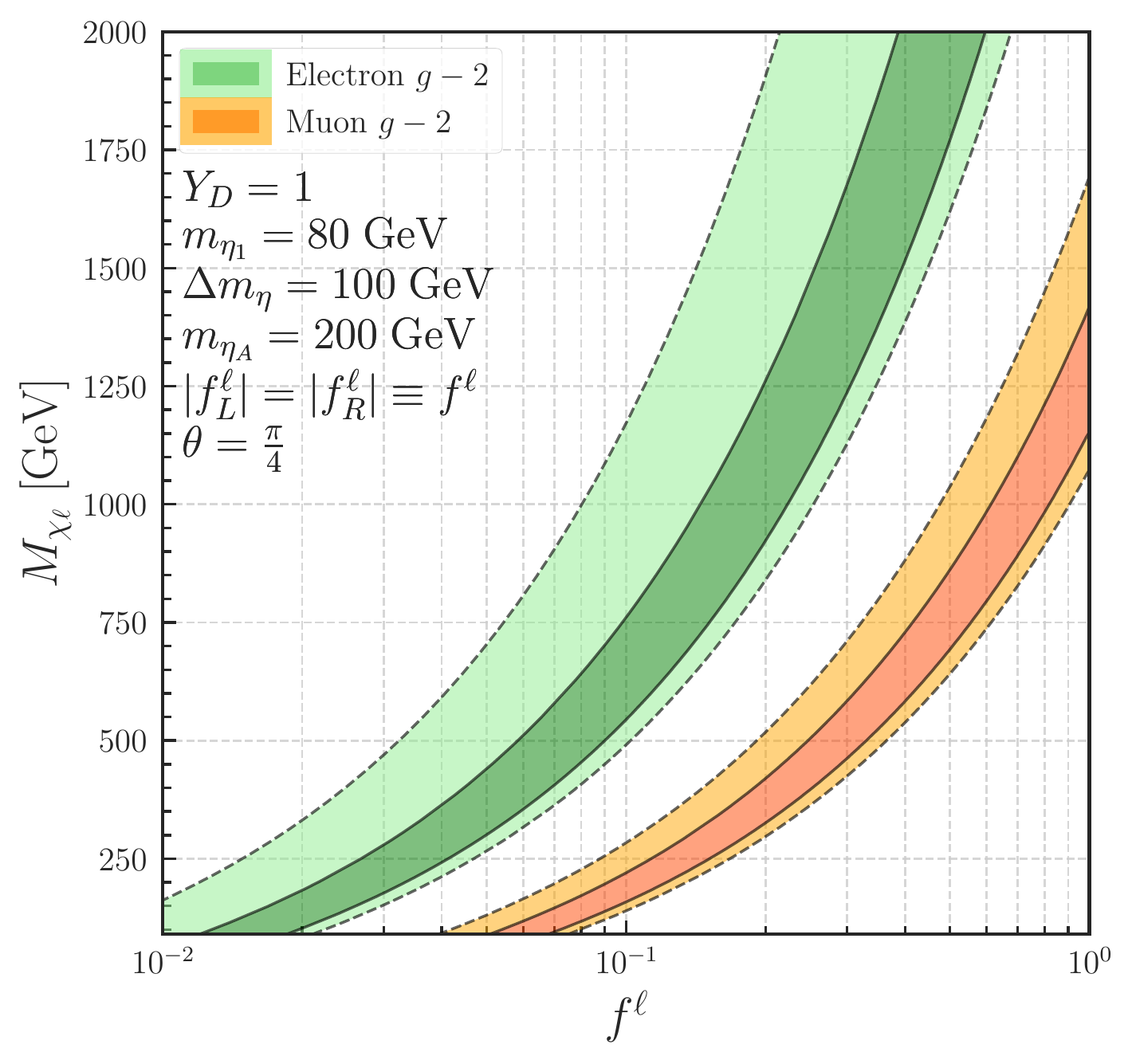}
    \hspace{5mm}
    \includegraphics[width=.45\textwidth]{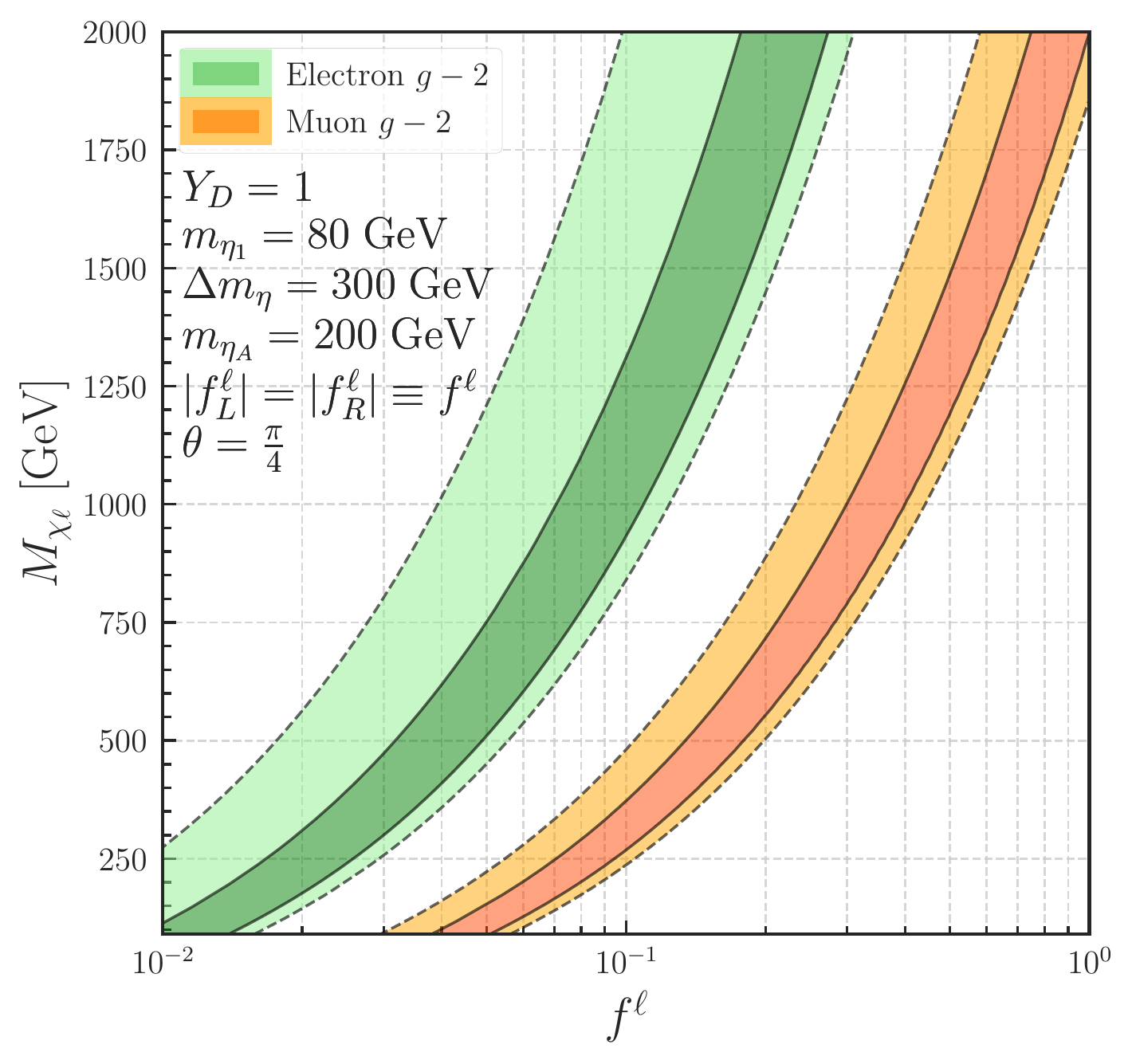}
    \caption{Regions in the plane of $f^\ell \equiv |f_L^\ell| = |f_R^\ell|$ and $M_{\chi_\ell}$ that can explain the corresponding $(g-2)_\ell$ for the scenario of $Y_D = 1$ at the $1\sigma$ (darker color) and $2\sigma$ (lighter color) levels.}
    \label{fig:g-2_Qx=1_mX-Yuk} 
\end{figure}

In Fig.~\ref{fig:g-2_Qx=1_mX-Yuk}, 
we show the regions in the plane of $f^\ell$ and the mass $M_{\chi_\ell}$ that can explain the corresponding $(g-2)_\ell$ anomalies in the scenario with $Y_D = 1$.  
The left and right panels show the allowed regions for a mass difference $\Delta m_{\eta} \equiv m_{\eta_2} - m_{\eta_1}$ of 100~GeV and 300~GeV, respectively.  
In this scenario, the lighter scalar $\eta_1^0$ can be the DM candidate and its mass $m_{\eta_1}$ is fixed to be 80~GeV.  
In the next section, we will see that this choice of the DM mass is compatible with both the observed relic density and the direct search experiments. 
It is clear that a smaller value of $\Delta m_{\eta}$ results in a larger cancellation between the $\Delta a^{\mathrm{NP}}_\ell$ contributions from the two scalar bosons, 
thus pushing the required Yukawa couplings higher for the same $M_{\chi_\ell}$.  Also, for a fixed $M_{\chi_\ell}$, the required value of $f^e$ is smaller than $f^\mu$ by roughly a factor of 4. 
This can be understood in such a way that from Eq.~(\ref{eq:da1}) the ratio $\Delta a_\mu^{\rm NP}/\Delta a_e^{\rm NP}$ is roughly given by $m_\mu/m_e \times |f^\mu/f^e|^2 \simeq 200 \times|f^\mu/f^e|^2$ 
if we take $M_{\chi_\mu} = M_{\chi_e}$. Therefore, with the required ratio $\Delta a_\mu/\Delta a_e$ by data to be about $3000$, 
the Yukawa coupling for the muon needed to explain the data should indeed be about 4 times larger than that for the electron.

\begin{figure}
    \centering
    \includegraphics[width=0.45\textwidth]{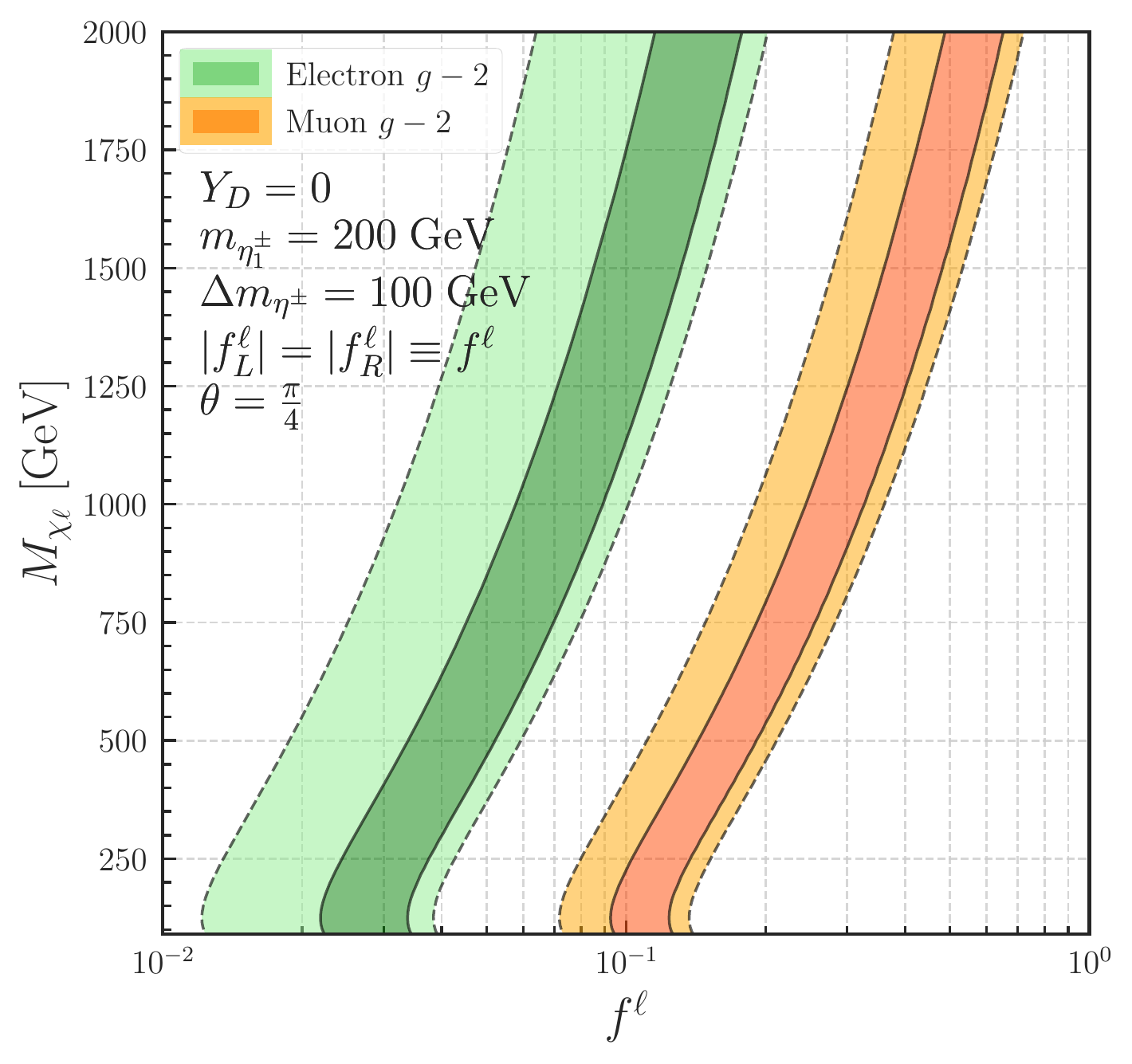}
	\hspace{5mm}
    \includegraphics[width=0.45\textwidth]{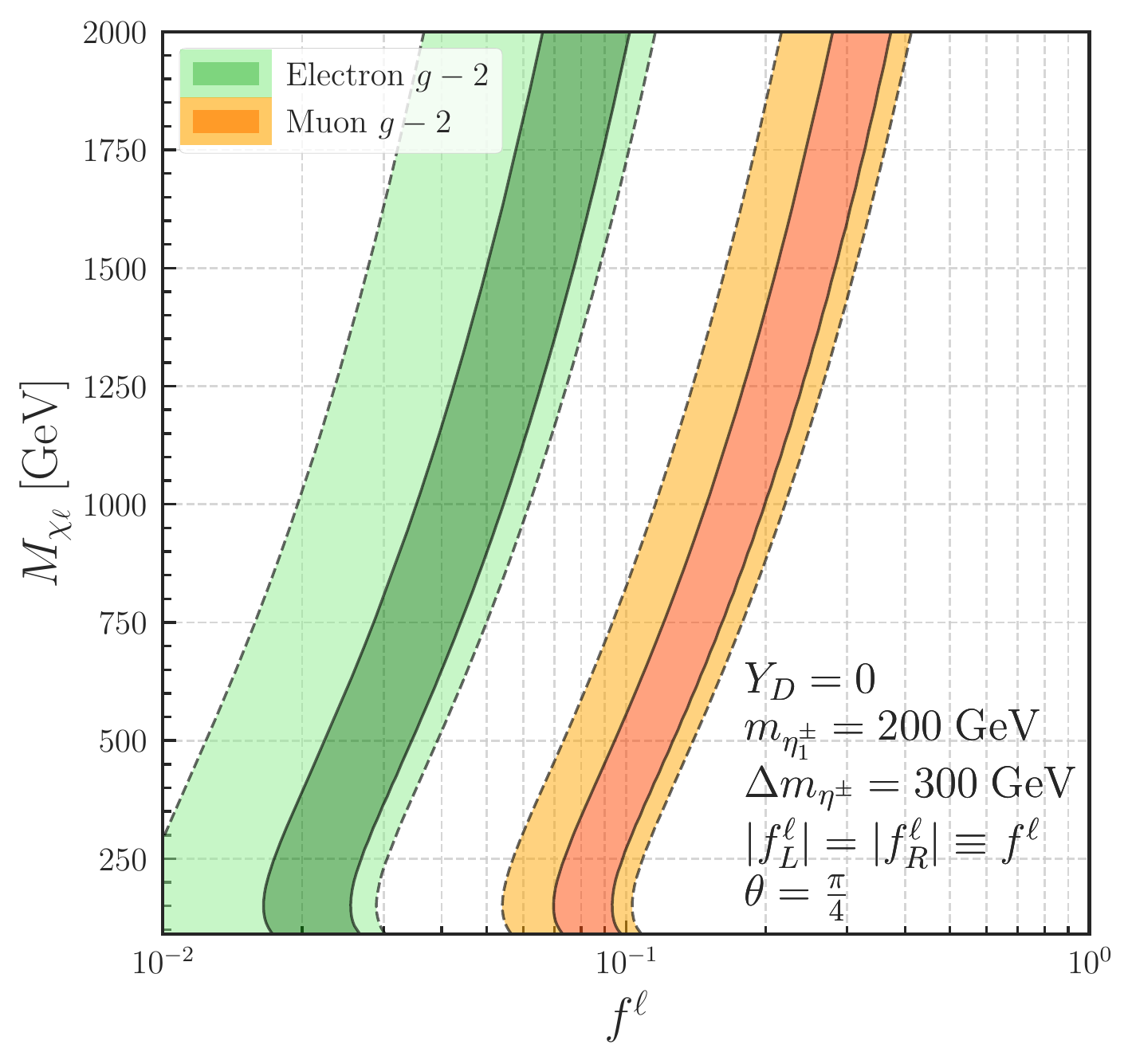}
    \caption{As in Fig.~\ref{fig:g-2_Qx=1_mX-Yuk}, but in the scenario of $Y_D = 0$.  The mass of the lighter charged scalar $\eta_1^\pm$ is set to be 200~GeV.} \label{fig:g-2_Qx=0_mX-Yuk}
\end{figure}

In Fig.~\ref{fig:g-2_Qx=0_mX-Yuk}, we show the results for $Y_D = 0$. 
In this scenario, the lighter charged scalar boson $\eta^\pm_1$ would not be a DM candidate and its mass $m_{\eta^\pm_1}$ would not be strongly constrained by the relic density and the direct search experiments.  
However, { $m_{\eta^\pm_1}$ of ${\cal O}(1)$~TeV} requires a large Yukawa coupling $f^\mu$ to explain the muon $g-2$ anomaly, which leads to too small a relic density to explain the observed density of DM as we will see 
in the next section. 
We thus take $m_{\eta^\pm_1}=200$~GeV as { a successful} example. 
In Fig.~\ref{fig:g-2_Qx=0_mX-Yuk}, 
we also observe a similar trend that for a fixed $M_{\chi_\ell}$, the required $f^e$ is smaller than $f^\mu$ by roughly a factor of 4 and both are pushed higher for smaller $\Delta m_{\eta^\pm}$. 
Unlike the scenario of $Y_D = 1$, the contours turn around at $M_{\chi_\ell} \sim 150$ GeV in this scenario. 
This is because the dominant term in Eq.~{\eqref{eq:da2}} reaches its maximum at $M_{\chi_\ell} = m_{\eta^\pm_k}$, so that 
the required value of $f^\ell$ becomes smallest at $M_{\chi_\ell} \sim 150$~GeV\footnote{For $Y_D = 1$, 
the dominant term in Eq.~({\ref{eq:da1}}) reaches its maximum at $M_{\chi_\ell} \sim 0.12 m_{\eta_1}$. 
Thus, the turning behavior is not observed as we take $\eta_1^0$ to be the lightest particle.}. 
Note that this turning point is lower in the left plot because of the larger cancellations for the case with $\Delta m_{\eta^\pm} = 100$~GeV (left) than that with $\Delta m_{\eta^\pm} = 300$~GeV (right).

We note that, in both scenarios with $Y_D = 1$ and $0$, the charged $\mathbb{Z}_2$-odd particles can be pair produced at colliders and their leptonic decays are subject to constraints from the experimental searches at the LHC. These constraints will be discussed in Sec.~\ref{subsec:collider}.

Lastly, we comment on the contributions from two-loop Barr-Zee type diagrams~\cite{Barr:1990vd}. 
In our model, new contributions to the Barr-Zee type diagrams can enter via the $\mathbb{Z}_2$-odd particle loops in the effective $h\gamma\gamma$, $hZ\gamma$ and $W^+ W^-\gamma$ vertices. The first two vertices, in particular, may give rise to sizable contributions to $\Delta a_{\ell}^{\rm NP}$, if the scalar trilinear couplings are taken to be large. 
However, such large values are highly constrained by the Higgs data to be discussed in Sec.~\ref{subsec:collider}. 
Together with the smallness of the Yukawa couplings for muon and electron, we find that contributions from these two types of diagrams are negligible. 
The contributions from diagrams with the $W^+ W^-\gamma$ effective vertex have been examined in detail in Ref.~\cite{Ilisie:2015tra}. 
It is shown that the contributions are at least two orders of magnitude smaller than the experimental measurements and can also be safely neglected.

\section{Phenomenology \label{sec:pheno}}

In this section, we discuss the phenomenological consequences of our models, focusing on the physics of DM and collider signatures of the new particles.

\subsection{Dark Matter Phenomenology\label{subsec:DM} }

\begin{figure}
    \centering
    \includegraphics[width=1.0\textwidth]{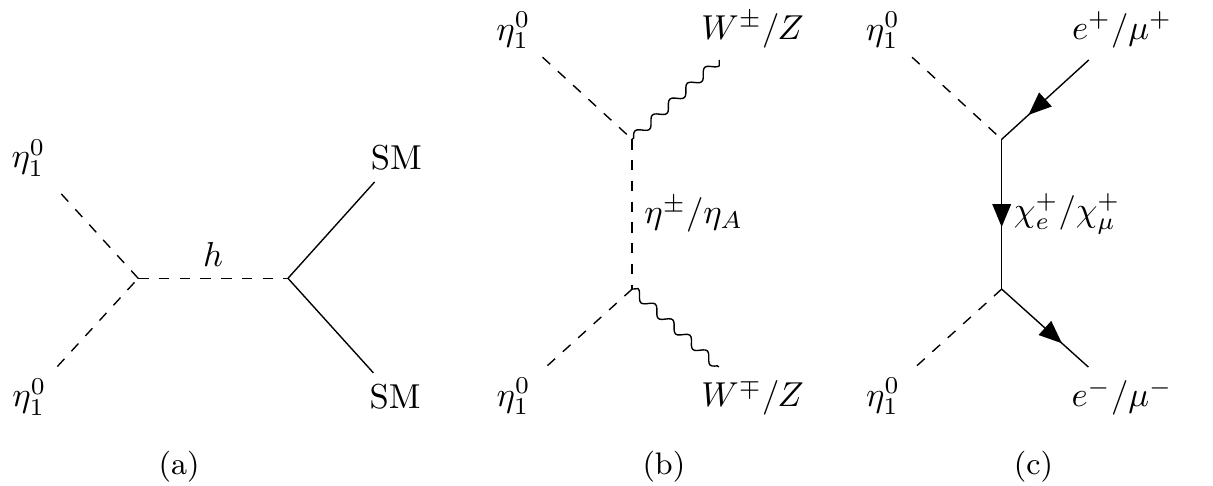}
    \caption{Important diagrams that contribute to the DM annihilation into the SM particles. }  
    \label{fig:DM_annihilation_case1}
\end{figure}

As alluded to in Sec.~\ref{sec:model}, the lightest neutral $\mathbb{Z}_2$-odd particle can be a DM candidate and corresponds to $\eta_1^0$ ($\eta_H^0$ or $\chi_\ell$) in the scenario of $Y_D=1$ ($Y_D=0$). 
Current measurements of the cosmic microwave background radiation by the Planck satellite show the DM relic density to be~\cite{Aghanim:2018eyx}
\begin{equation}
    \Omega_{\mathrm{DM}}h^2 = 0.120 \pm 0.001,
\end{equation}
assuming the cold DM scenario. 

We first discuss the relic density of DM in the scenario of $Y_D = 1$.  The important DM annihilation processes are shown in Fig.~\ref{fig:DM_annihilation_case1}.  
The amplitude of the $s$-channel Higgs-mediated process is proportional to the $\eta_1^0\eta_1^0 h$ coupling calculated as
\begin{equation}
\lambda_{\eta_1^0 \eta_1^0 h} = v\left[c_{\theta}^2 \left(\frac{m_{\eta^{\pm}}^2}{v^2} - \frac{m_{\eta_{1}}^{2}}{v^2} -\frac{\lambda_{3}}{2} \right)-\lambda_7s_{\theta}^2\right],  \label{eq:h11}
\end{equation}
where the $\lambda_3$ and $\lambda_7$ parameters are chosen as independent parameters [see Eqs.~\eqref{eq:free-para-YD1} and \eqref{eq:free-para-YD0}] in our analyses. 
Therefore, the $\lambda_{\eta_1^0 \eta_1^0 h}$ coupling can be taken to be any value as far as it satisfies the theoretical bounds discussed in Sec.~\ref{sec:model}.  
This process can be particularly important when the DM mass is close to half of the Higgs boson mass due to the resonance effect.  
The amplitude of the $t$-channel process mediated by the heavier $\mathbb{Z}_2$-odd scalar bosons becomes important when the DM mass is larger than about 80~GeV because of the threshold of the weak gauge boson channels.  
The $t$-channel process mediated by the vector-like lepton $\chi_\ell$ is sensitive to the Yukawa couplings $f_{L,R}^\ell$, while weakly depending on the mass of the lighter vector-like lepton.  
In addition to the processes shown in Fig.~\ref{fig:DM_annihilation_case1}, we also take into account the contributions from DM co-annihilations with the heavier $\mathbb{Z}_2$-odd particles, {\it i.e.}, $\eta_A^{0}$, $\eta_2^0$, $\eta^{\pm}$ and $\chi^{\pm}_{\ell}$. 
For numerical calculations, we have implemented our model using \texttt{FeynRules}~\cite{Alloul:2013bka, Degrande:2011ua} and derived the relic density and direct search constraints using \texttt{MadDM}~\cite{Ambrogi:2018jqj, Backovic:2015cra, Backovic:2013dpa}.

\begin{figure}
    \centering
    \includegraphics[width=1.0\textwidth]{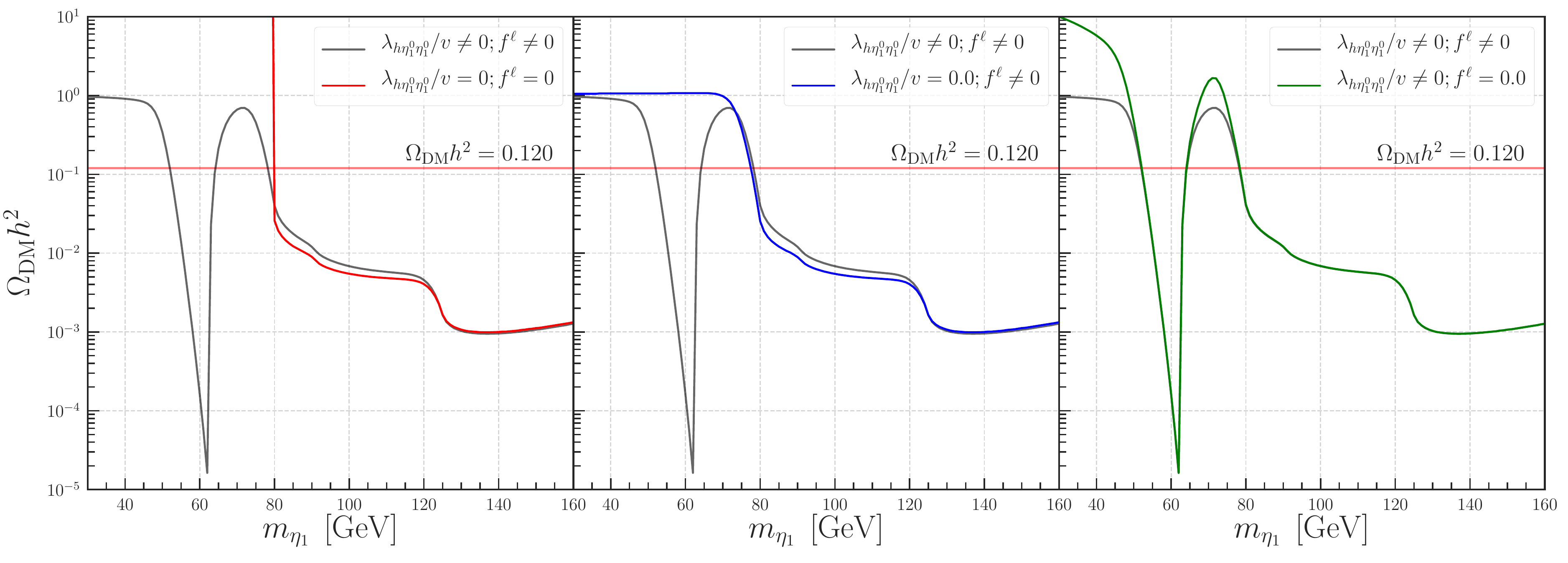}
    \caption{Contributions of different processes shown in Fig.~\ref{fig:DM_annihilation_case1} to the DM relic density in the model with $Y_D=1$ as a function of the DM mass $m_{\eta_1}$. 
The grey curves show the case for the benchmark parameter set with the mass spectrum $(m_{\eta_2}, m_{\eta_A}, m_{\eta^\pm}, M_{\chi_e}, M_{\chi_\mu}) = (380, 200, 200, 1100, 600)$~GeV and the coupling strengths  
$(f^e, f^\mu, \lambda_{h\eta^0_1\eta^0_1}/v) = (0.1, 0.2, 0.01)$. 
From the left to right panels, the colored curve shows the case with some of the couplings taken to be zero, by which we see the impact of the contribution
from the process of (b), (b) plus (c) and (a) plus (b) shown in Fig.~\ref{fig:DM_annihilation_case1}.     \label{fig:DM_Signature} }
\end{figure}

Fig.~\ref{fig:DM_Signature} shows a typical behavior of the DM relic density as a function of the DM mass $m_{\eta_1}$ in the model with $Y_D= 1$. In all three panels, 
the grey curves show a benchmark case with the parameter choice $(f^e, f^\mu, \lambda_{h\eta^0_1\eta^0_1}/v) = (0.1, 0.2, 0.01)$ 
and $(m_{\eta_2}, m_{\eta_A}, m_{\eta^\pm}, M_{\chi_e}, M_{\chi_\mu}) = (380, 200, 200, 1100, 600)$~GeV, where 
$M_{\chi_\ell}$ are determined according to Fig.~\ref{fig:g-2_Qx=1_mX-Yuk} such that both electron and muon $g-2$ anomalies can be accommodated within $1\sigma$ at $m_{\eta_1} = 80~\mathrm{GeV}$.
By turning off some of the couplings, we show with colored curves in the three panels how the relic density changes if only a subset of the processes in Fig.~\ref{fig:DM_annihilation_case1} is taken into account.  
The leftmost plot of Fig.~\ref{fig:DM_Signature} shows that for $m_{\eta_1} \gtrsim 80~\mathrm{GeV}$, the $t$-channel annihilations into weak gauge bosons are kinematically allowed and become the dominant process.  
It is clear from the central plot that for $m_{\eta_1} < 50~\mathrm{GeV}$, the relic density is dominated by the $t$-channel annihilations into electron and muon pairs.   The rightmost plot shows that the Higgs-mediated $s$-channel process is most important around the Higgs resonance when $m_{\eta_1}\sim 62.6~\mathrm{GeV}$.  We observe that for $m_{\eta_1} < 150~\mathrm{GeV}$, there are three solutions to the relic density: 
one at $m_{\eta_1}\sim 80~\mathrm{GeV}$ and the remaining two around half the Higgs resonance.

\begin{figure}
    \centering
\includegraphics[width=1.0\textwidth]{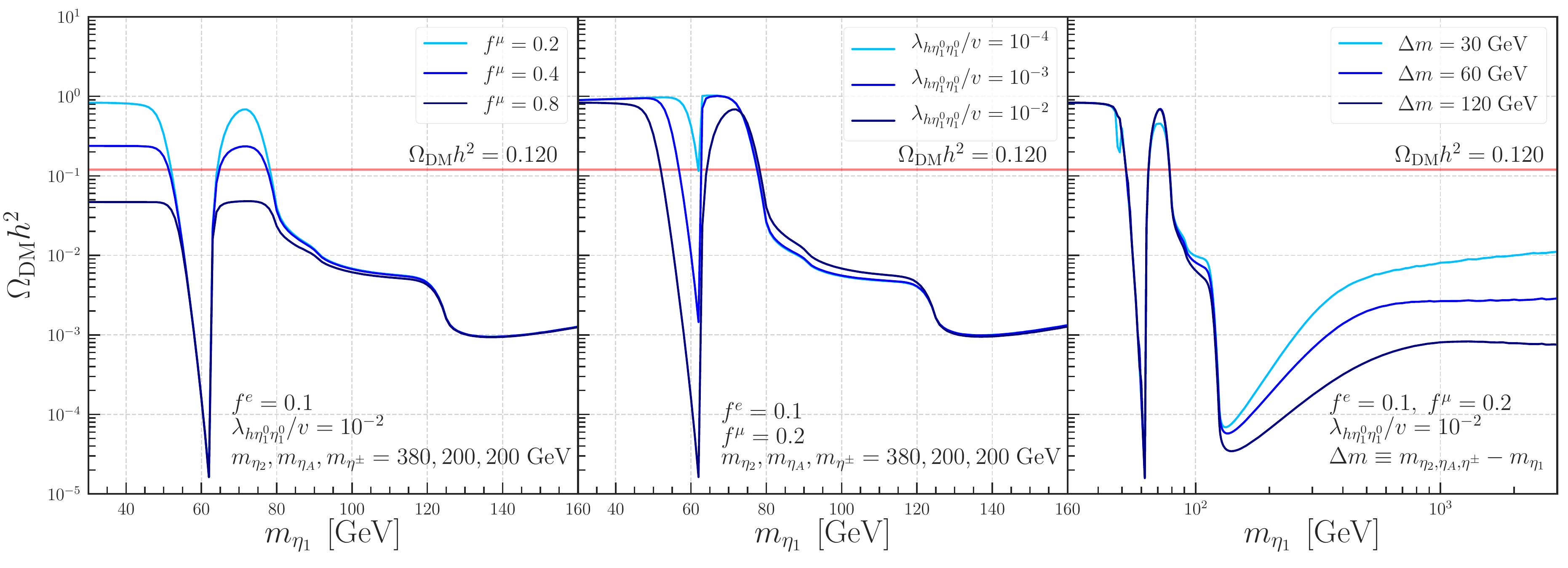}
    \caption{Relic density as a function of the DM mass $m_{\eta_1}$ in the model with $Y_D = 1$. 
The left, center and right panel shows, respectively, the effect of varying the magnitude of the Yukawa coupling $f^\mu$, the $\lambda_{h\eta^0_1\eta^0_1}$ coupling, and the mass splitting $\Delta m$ 
(with $m_{\eta_2} = m_{\eta_A} = m_{\eta^\pm}$) defined in the figure.
For all the { panels}, $M_{\chi_e}-m_{\eta_1}$ is fixed to be $1020$ GeV, while 
$M_{\chi_\mu}-m_{\eta_1}$ is taken to be $520$ (1070) [1820] GeV for $f^\mu = 0.2$ (0.4) [0.8]
such that the $g-2$ anomalies can be explained within $1\sigma$ level, where the latter two choices are only taken in the left plot. }
    \label{fig:DM_param} 
\end{figure}

The impacts of the key parameters in each process shown in Fig.~\ref{fig:DM_annihilation_case1} are investigated in Fig.~\ref{fig:DM_param}. 
From left to right, we investigate the dependence on the magnitude of the Yukawa coupling $f^\mu$, the $\lambda_{h\eta^0_1\eta^0_1}$ coupling, and the mass splitting $\Delta m$ between the DM 
and all the other heavier $\mathbb{Z}_2$-odd scalar bosons. 
From the left two plots, we see that an increase in $f^\ell$ reduces the overall relic density in the low-mass region 
while a decrease in $\lambda_{h\eta^0_1\eta^0_1}$ makes the dip around the Higgs resonance shallower. 
In the leftmost (center) plot, we find the critical values $f^\mu \simeq 0.54$ \footnote{A more conservative upper limit for 
the magnitude of the Yukawa coupling is found to be $0.34$ for the case with $f^e = f^\mu$.}  
$(\lambda_{h\eta_1^0\eta_1^0}/v \simeq 10^{-4})$ above (below) which the solutions of $m_{\eta_1}$ to realize the observed relic density disappears.
In addition, if we take $\lambda_{h\eta^0_1\eta^0_1}/v \gtrsim 0.10$ in the center plot, 
the solutions at $m_{\eta_1} \geq  m_h/2$ disappear because the dip becomes too deep.

It is worth mentioning that in the Inert Doublet Model (IDM), 
another solution of the DM mass to satisfy the relic density may exist in a TeV region when the mass splitting among the $\mathbb{Z}_2$-odd scalar particles is small, typically less than 10 GeV \cite{LopezHonorez:2006gr}. 
In such a scenario, DM dominantly annihilates into a pair of weak gauge bosons whose annihilation cross section decreases by ${\cal O}(1/m^2_{\mathrm{DM}})$, while 
the annihilation into the Higgs bosons is highly suppressed due to small Higgs--DM couplings. 
In our model, such a high mass solution cannot be realized, because the additional $\eta_2^0$ state cannot have the mass close to $\eta_1^0$ in order to explain the $g-2$ anomaly as discussed in Sec.~{\ref{sec:g-2}}. 
As a result, the (co)annihilation into a pair of the Higgs bosons is not suppressed at the high mass region. 
This situation can be clearly seen in the right panel of Fig.~{\ref{fig:DM_param}} in which we take $\Delta m = 30,~60,~120~\mathrm{GeV}$ that can explain the $g-2$ anomalies.
Indeed, the predicted density is well below the observed value at the high mass region. 
In fact, we confirm that solutions do not appear even at a few hundred TeV of $m_{\eta_1}$. 

\begin{figure}
    \centering
    \includegraphics[width=0.7\textwidth]{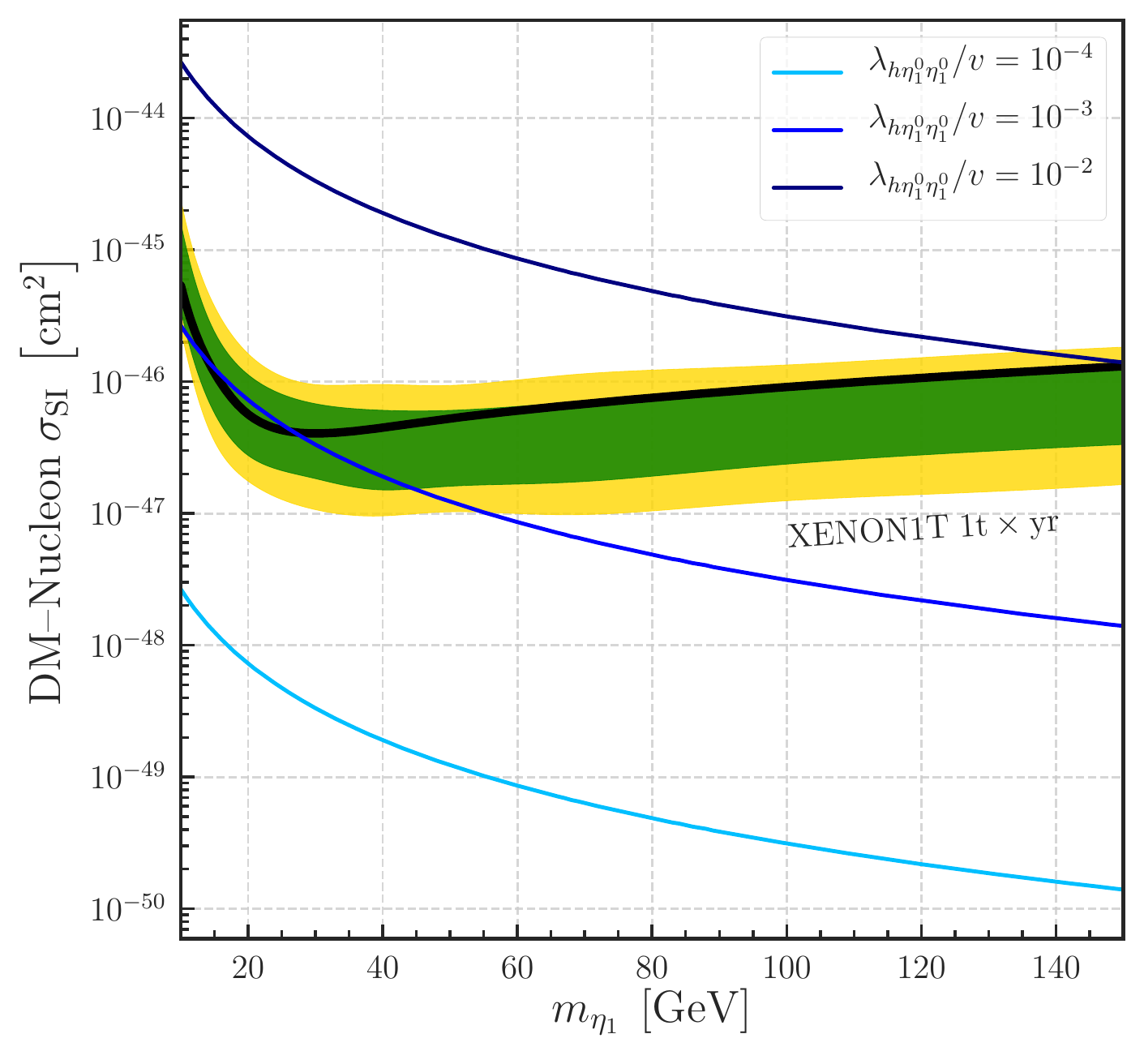}
    \caption{Spin-independent DM--Nucleon scattering cross section as a function of the DM mass $m_{\eta_1}$ for several values of the $\lambda_{h\eta^0_1\eta^0_1}$ coupling. 
The black curve shows the 90\% confidence level upper limit obtained from the XENON1T experiment with a $1.0\textrm{t}\times 1\mathrm{yr}$ exposure. 
The green and yellow region marks the $1$ and $2\sigma$ sensitivity bands for the XENON1T results. \label{fig:DM_direct} }
\end{figure}

In addition to the DM annihilation, the $\lambda_{h\eta^0_1\eta^0_1}$ coupling contributes to the scattering of DM with nuclei via the mediation of the Higgs boson, 
allowing our DM candidate to be probed by the direct search experiments. 
Fig.~\ref{fig:DM_direct} shows the spin-independent DM--nucleon scattering cross section and 
its upper limit at 90\% confidence level obtained from the XENON1T experiment with a 1-tonne times one year exposure~\cite{Aprile:2018dbl}. 
We find that $\lambda_{h\eta^0_1\eta^0_1}/v$ has to be smaller than 0.0026, 0.0034, and 0.0047 for the DM $\eta_1^0$ to { have} a mass around $50$, $65$ and $80$~GeV, respectively, by which we can explain the observed relic density. 

In conclusion, the mass of $\eta_1^0$ should be about $50$, $65$ or $80$~GeV while having $f^\ell \lesssim 0.34$ and $\lambda_{h\eta_1^0\eta_1^0}/v \in \left[1.0 \times 10^{-4}, 2.6 \times10^{-3}\right]$ 
in order to satisfy both the relic density and the direct search experiment in the scenario with $Y_D = 1$.

Next, we discuss the scenario with $Y_D=0$ assuming $\eta_H^0$ to be the DM candidate. 
In this scenario, the properties of DM are quite similar to those of the scenario with $Y_D=1$ discussed above, where 
the annihilation processes can be obtained by replacing ($\eta_1^0$,$\eta^\pm$,$e/\mu$) with ($\eta_H^0$,$\eta_{1,2}^\pm$,$\nu_e/\nu_\mu$) in Fig.~\ref{fig:DM_annihilation_case1}. The $\eta_H^0\eta_H^0h$ coupling is given as
\begin{equation}\label{eq:h00}
 \lambda_{\eta_H^0 \eta_H^0 h} = \frac{v}{2}\left(\frac{m^2_{\eta_A}}{v^2} - \frac{m^2_{\eta_H}}{v^2} - \lambda_3\right). 
\end{equation}
Again, this coupling can be taken to be any value due to the independent parameter $\lambda_3$ as far as it satisfies the theoretical constraints. Taking similar values of the Higgs to DM coupling and the new Yukawa couplings as those in the model with $Y_D=1$, we obtain almost identical results as in Figs.~\ref{fig:DM_Signature} and ~\ref{fig:DM_param}, with minor modifications due to the changes in $M_{\chi_\ell}$ in order to satisfy the $(g-2)_\ell$ anomalies. 

Finally, we briefly comment on the other possibility of having $\chi_\ell$ as the DM candidate in the model with $Y_D = 0$. 
The dominant annihilation channels for $\chi_\ell$ are the $t$-channel processes $\chi_\ell \bar{\chi}_\ell \to \nu_\ell \bar{\nu}_\ell/\ell^+\ell^-$ mediated by a neutral or charged $\mathbb{Z}_2$-odd scalar boson. These processes alone, however, produce a cross section that is too small to account for the observed relic density. Thus, the scenario of having a fermionic DM in our model is ruled out.

\subsection{Collider Phenomenology \label{subsec:collider}}

We first discuss the constraints from direct searches for new particles at high-energy collider experiments.  
In our model, all the new particles are $\mathbb{Z}_2$-odd, and thus would only be produced in pairs at colliders.  In addition, due to the new Yukawa interactions for the muon and the electron, their decays typically include a muon or an electron in association with missing energy carried away by the DM.   Therefore, our model can be tested by looking for an excess of events with multiple charged leptons plus missing energy, which is identical to the signatures of slepton or chargino production in supersymmetric models. 

\begin{figure}
    \centering
    \includegraphics[width=1.0\textwidth]{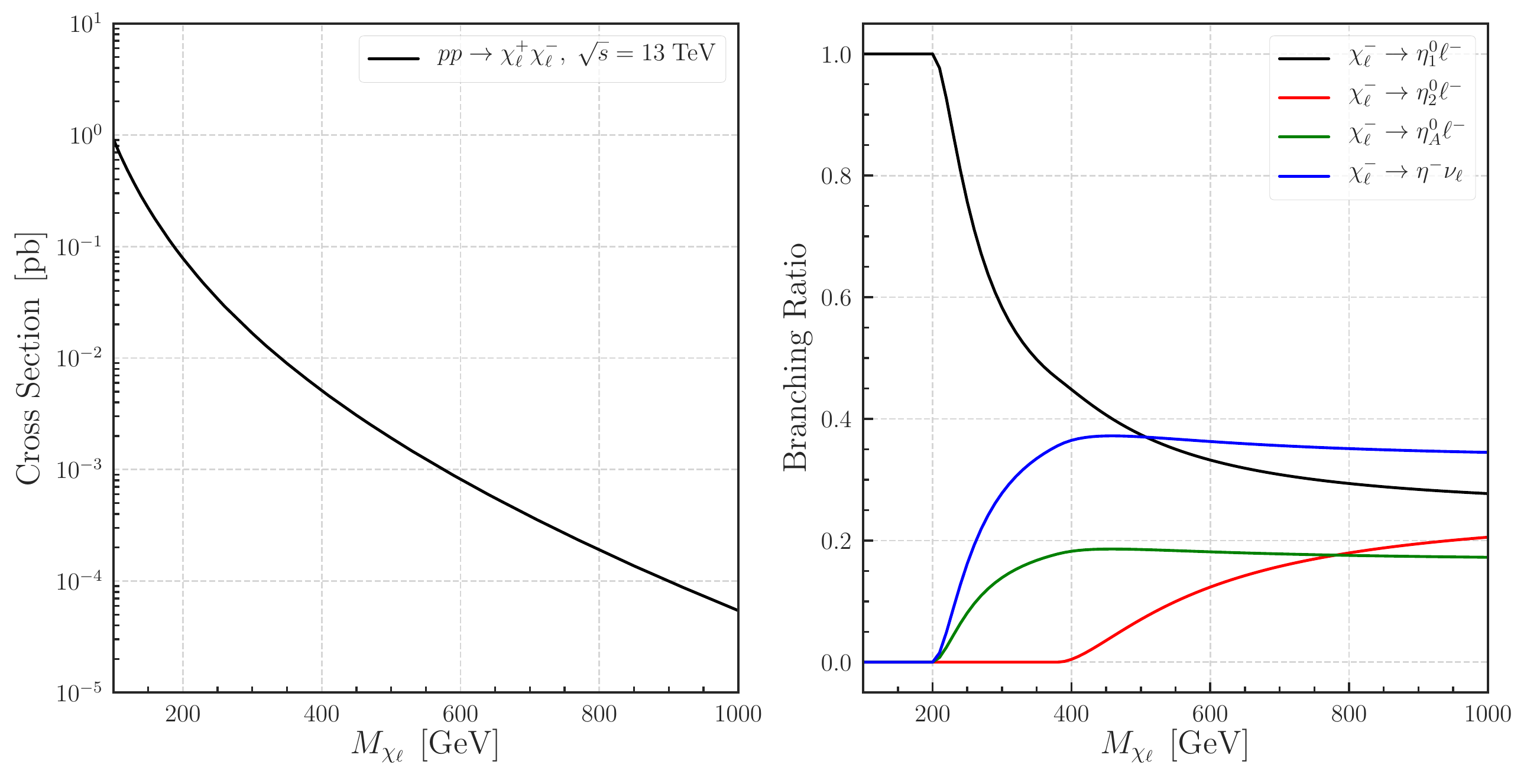}
    \caption{\textit{Left:} Cross section of $p p \rightarrow \chi_\ell^+ \chi_\ell^-$ as a function of $M_{\chi_\ell}$ in the model with $Y_D = 1$ at $\sqrt{s} = 13~\mathrm{TeV}$. 
\textit{Right:} Branching ratios of $\chi_\ell$ in the model with $Y_D = 1$ with ($m_{\eta_1}$, $m_{\eta_A}$, $m_{\eta^\pm}$, $m_{\eta_2}$) = (80, 200, 200, 380)~GeV and $\theta = \pi/4$. \label{fig:xsec_br_chi} }
\end{figure}

We first focus on the pair production of the vector-like leptons $\chi_\ell^\pm$ at the LHC in the model with $Y_D = 1$.  
The pair production occurs via the Drell-Yan process mediated by the photon and $Z$ boson, so that its cross section is simply determined by the mass of $\chi_\ell$.  
The left panel of Fig.~\ref{fig:xsec_br_chi} shows the cross section of $pp \to \gamma^*/Z^* \to \chi_\ell^+\chi_\ell^-$ with the collision energy of 13~TeV. The cross section is calculated at the leading order using \texttt{MadGraph\char`_aMC@NLO} \cite{Alwall:2014hca} with the parton distribution functions \texttt{NNPDF23\char`_lo\char`_as\char`_0130\char`_qed} \cite{Ball:2013hta}. It is seen that the cross section is about 900, 20 and 0.8~fb for $M_{\chi_\ell}=150$, 300 and 600~GeV, respectively. 
On the other hand, the decays of $\chi_\ell^\pm$ strongly depend on the mass spectrum of the $\mathbb{Z}_2$-odd scalar bosons.  
For the case with ($m_{\eta_1^0}$, $m_{\eta_A}$, $m_{\eta^\pm}$, $m_{\eta_2^0}$) = (80, 200, 200, 380)~GeV, the various decay branching ratios of $\chi_\ell^\pm$ 
are depicted in the right panel of Fig.~\ref{fig:xsec_br_chi}.  In this plot, we take $\theta = \pi/4$ in which the branching ratios do not depend on $f^\ell$. 
We see that $\chi_\ell^\pm$ decay $100\%$ into $\eta_1^0\ell^\pm$ when $M_{\chi_\ell} < 200 ~\mathrm{GeV}$ because this is the only kinematically allowed channel.  
At higher masses, $\chi_\ell^\pm$ can also decay into $\eta_2^0\ell^\pm$, $\eta_A^0\ell^\pm$ and $\eta^\pm \nu_\ell$.  The heavier $\mathbb{Z}_2$-odd scalar bosons can further decay into the DM and a SM particle, {\it i.e.}, $\eta_2^0\to h\eta_1^0$, $\eta_A^0\to Z\eta_1^0$, and $\eta^\pm \to W^\pm\eta_1^0$. Therefore, when these channels are allowed, the final state of the $\chi_\ell^\pm$ decays can have 1 or 3 charged leptons.  We note that the tri-lepton channel is highly suppressed by the small branching ratio of the leptonic decays of the $Z$ boson or the Higgs boson.

\begin{figure}
    \centering
    \includegraphics[width=0.7\textwidth]{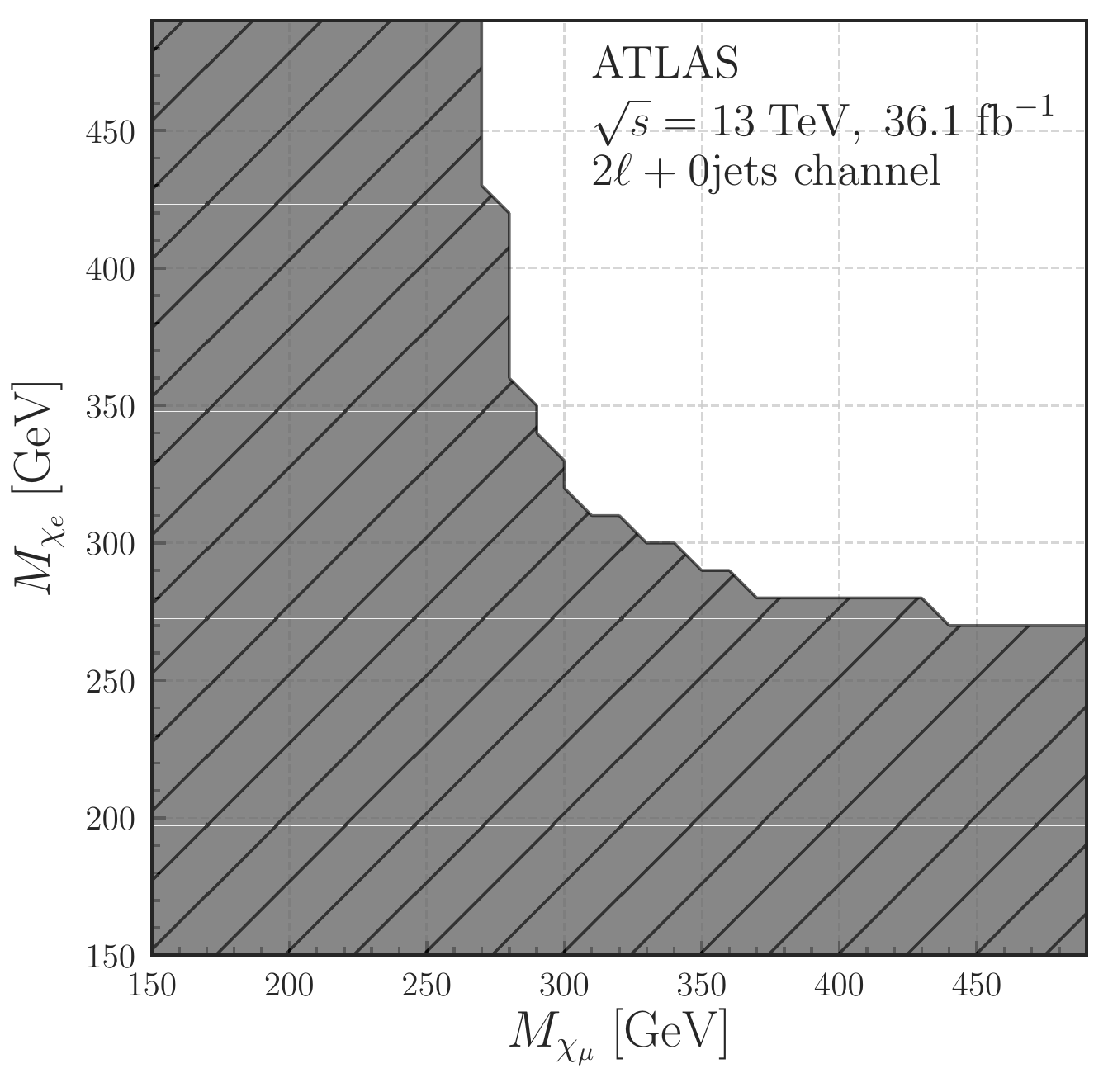}
    \caption{Excluded region in the plane of the masses of vector-like leptons $M_{\chi_\mu}$--$M_{\chi_e}$ in the model with $Y_D = 1$
from the searches for events with exactly two or three electrons or muons and missing transverse momentum by the ATLAS experiment with $\sqrt{s}=13~\mathrm{TeV}$ and $36.1~\textrm{fb}^{-1}$ of the integrated luminosity. 
We take $(m_{\eta_1}, m_{\eta_2}, m_{\eta_A}, m_{\eta^\pm}, \theta) = (80~\mathrm{GeV}, 380~\mathrm{GeV}, 200~\mathrm{GeV}, 200~\mathrm{GeV}, \pi/4)$. }
\label{fig:chi_excluded-ATLAS13TeV}
\end{figure}

In Fig.~\ref{fig:chi_excluded-ATLAS13TeV}, we show the observed exclusion limit on the vector-like lepton masses $M_{\chi_\ell}$ using the same set of parameters as in Fig.~\ref{fig:xsec_br_chi}.  
The observed limit is derived based on the searches for events with exactly two or three electrons or muons and missing transverse momentum performed by the ATLAS experiment 
using the $36.1~\textrm{fb}^{-1}$ dataset of $\sqrt{s}=13~\mathrm{TeV}$ collisions~\cite{Aaboud:2018jiw}. 
We use \texttt{MadGraph\char`_aMC@NLO} \cite{Alwall:2014hca} to simulate the events and to compute the $\chi^+_\ell\chi^-_\ell$ production cross section at the leading order. 
The events are further processed by \texttt{Checkmate}~\cite{Dercks:2016npn, Cacciari:2008gp, Cacciari:2011ma, Read:2002hq}, which utilizes \texttt{Pythia8}~\cite{Sjostrand:2006za, Sjostrand:2007gs} for parton showering and hadronization and \texttt{Delphes3}~\cite{deFavereau:2013fsa} for detector simulations and compares the number of events with the limit in a given signal region provided by the ATLAS experiment~\cite{ATLAS:2016uwq}.  With our parameter choice, $M_{\chi_\ell} \alt 270$~GeV is excluded.  Note also that such lower bounds on the $\chi_\ell$ mass depend on the mass spectrum of the $\mathbb{Z}_2$-odd scalar bosons, 
and are usually lower than the bounds extracted in the literature.
{For example, the branching ratio of $\chi_\ell^\pm \to \eta_1 \ell^\pm$ is assumed to be 100\% in Ref.~\cite{Calibbi:2018rzv}, while we take other decay channels (see Fig.~\ref{fig:xsec_br_chi}) into account as well and thus obtain a less stringent constraint.}

\begin{figure}
    \centering
    \includegraphics[width=0.7\textwidth]{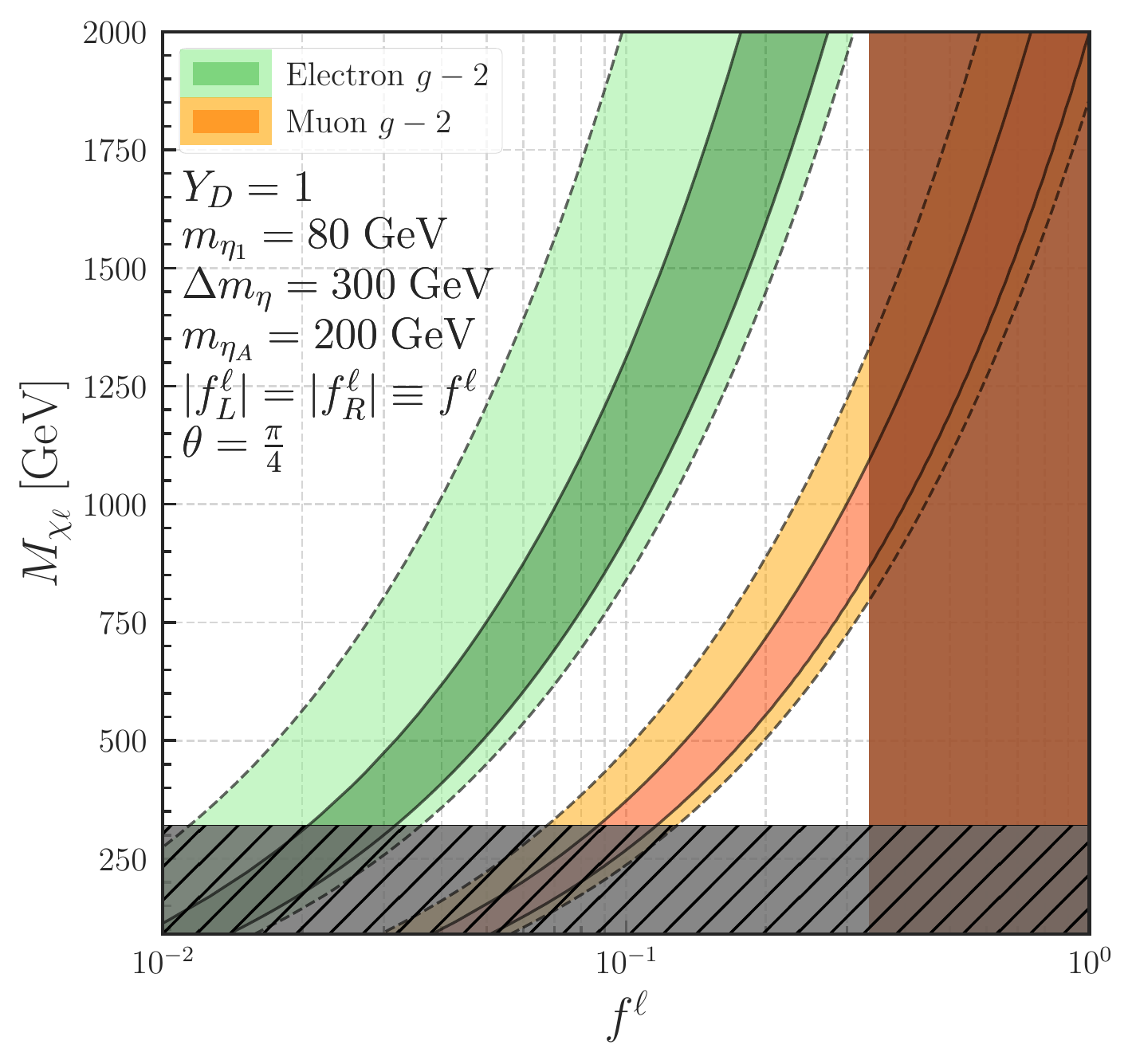}
    \caption{Summary of the constraints in the plane of $f^\ell$ and $M_{\chi_\ell}$ for the benchmark case with $Y_D = 1$ and  
$(m_{\eta_1}, m_{\eta_2}, m_{\eta_A}, m_{\eta^\pm}, \theta) = (80~\mathrm{GeV}, 380~\mathrm{GeV}, 200~\mathrm{GeV}, 200~\mathrm{GeV}, \pi/4)$. The regions shaded by dark green and orange can explain the electron and the muon $g-2$ within $1\sigma$.  The lower bounds on $M_{\chi_\ell}$ are derived from the direct search limit by the ATLAS collaboration,  
while the brown area cannot explain the observed DM relic density.} 
\label{fig:Y_D=1-Summary}
\end{figure}

In Fig.~\ref{fig:Y_D=1-Summary}, we summarize all the constraints discussed above in our model with $Y_D = 1$.  
The regions shaded by dark green and orange can explain, respectively, the electron and muon $g-2$ within $1\sigma$. 
The lower bound on $M_{\chi_\ell}$ is derived from the observed direct search limit by the ATLAS collaboration (see Fig.~\ref{fig:chi_excluded-ATLAS13TeV}), 
while the region shaded by brown cannot explain the DM relic density as the annihilation cross section of DM in this region is too large to reach the observed density (see Fig.~\ref{fig:DM_param}). 

We note that in addition to the pair production of $\chi_\ell^\pm$, the inert scalar bosons can also be produced in pairs. 
When we consider the case where the vector-like lepton masses are larger than the masses of the inert scalar bosons, 
the signature of these scalar bosons become quite similar to that given in the IDM. 
As shown in Ref.~\cite{Dercks:2018wch}, the upper limit on the cross section of multi-lepton final states given by the LHC Run-II data is 
typically one or more than one order of magnitude larger than that predicted in the IDM. 
Thus, we can safely avoid the bound from the direct searches for the inert scalar bosons at the LHC.

Let us briefly comment on the collider signatures in the model with $Y_D = 0$. 
In this scenario, the vector-like lepton is electrically neutral, so that it is not produced in pair via the Drell-Yan process, but can be produced from decays of the inert scalar bosons, e.g., 
$\eta_{1,2}^\pm \to \ell^\pm \chi_\ell^0$ and $\eta_{H,A}^0 \to \nu_\ell\chi_\ell^0$. 
The most promising process to test this scenario could then be a pair production of the charged inert scalar bosons $pp \to \eta_i^\pm \eta_j^\mp$ ($i,j=1,2$).  
However, we find that the production cross sections of $\eta^\pm_{1,2}$ are roughly one order of magnitude smaller than those of vector-like leptons shown in Fig.~\ref{fig:xsec_br_chi}, so that 
such process is more weakly constrained by the current LHC data as compared with that in the model with $Y_D = 1$. 

Finally, we discuss an indirect test of our model by focusing on modifications in the Higgs boson couplings. Because of the $\mathbb{Z}_2$ symmetry, the Higgs boson couplings do not change from their SM values at tree level.
However, the loop-induced $h\gamma\gamma$ and $hZ\gamma$ couplings can be modified due to the new charged scalar boson loops, {\it i.e.}, $\eta^\pm$ ($\eta_1^\pm$ and $\eta_2^\pm$)  in the model with $Y_D = 1$ ($Y_D = 0$). 
In order to discuss the modifications to the $h \to \gamma\gamma$ and $h \to Z\gamma$ decays, we introduce the signal strength $\mu_{\gamma\gamma}$ and $\mu_{Z\gamma}$ defined as follows: 
\begin{equation}
 \mu_{\gamma\gamma/Z\gamma} \equiv  \frac{\sigma_h \times \mathrm{BR}(h \to \gamma\gamma/Z\gamma) }{[\sigma_h \times \mathrm{BR}(h \to \gamma\gamma/Z\gamma)]_{\mathrm{SM}}}.  
\end{equation}
In our model, the production cross section of the Higgs boson should be the same as in the SM.  
Consequently, these signal strengths are simply given by the ratio of the branching ratio between our model and the SM. 
The decay rates of $h \to \gamma\gamma$ and $h \to Z\gamma$ depend on the Higgs boson couplings to the charged scalar bosons, which are calculated as
\begin{equation}
   \label{eq:hS+S-_YD=1}
   \lambda_{h\eta^+\eta^-} = -v\lambda_3   \quad   \text{for } Y_D = 1,
\end{equation}
and for $Y_D$ = 0,
\begin{equation}
\label{eq:hS+S-_YD=0}
\lambda_{h\eta_k^+\eta_k^-} = 
   \begin{cases}
    v\displaystyle\left[\left(\frac{m_{\eta_A}^2}{v^2} + \frac{m_{\eta_H}^2}{v^2} -\frac{2m_{\eta_1^\pm}^2}{v^2} - \lambda_3 \right)c_\theta^2 - \lambda_7s_\theta^2\right] \quad \textrm{for } k = 1,\\
   \\ 
   v\displaystyle\left[\left(\frac{m_{\eta_A}^2}{v^2} + \frac{m_{\eta_H}^2}{v^2} -\frac{2m_{\eta_2^\pm}^2}{v^2} - \lambda_3 \right)s_\theta^2 - \lambda_7c_\theta^2\right] 
   \quad \textrm{for } k = 2.
   \end{cases} 
\end{equation}
We note that the parameter $\lambda_3$ in the model with $Y_D = 1$ also appears in the DM coupling $\lambda_{h\eta_1^0\eta_1^0}$ [see Eq.~(\ref{eq:h11})], but the dependence of $\lambda_{h\eta_1^0\eta_1^0}$ on $\lambda_7$ makes it still possible to choose $\lambda_3$ freely. For the model with $Y_D = 0$,  $\lambda_3$ is controlled by the DM coupling $\lambda_{h\eta_1\eta_1}$ [see Eq.~(\ref{eq:h00})], but the $\lambda_{h\eta_k^+\eta_k^-}$ couplings can be chosen freely due to their dependence on the $\lambda_7$ parameter. In both scenarios, the new fermions $\chi_\ell$ do not couple to the Higgs boson as they are vector-like.

The current global average of the Higgs diphoton signal strength is given by $\mu_{\gamma\gamma}^{\rm Exp} = 1.10^{+0.10}_{-0.09}$~\cite{Tanabashi:2018oca}, 
where the deviation of the central value from the SM expectation mainly originates from the CMS measurements~\cite{Sirunyan:2018ouh}. 
On the other hand, the $h \to Z\gamma$ decay has not yet been observed, and the strongest limit is given by the ATLAS experiment~\cite{Aaboud:2017uhw}, where the observed upper limit for the signal strength $\mu_{Z\gamma}$ is $6.6$ at 95\% confidence level.

\begin{figure}
    \centering
    \includegraphics[width=0.45\textwidth]{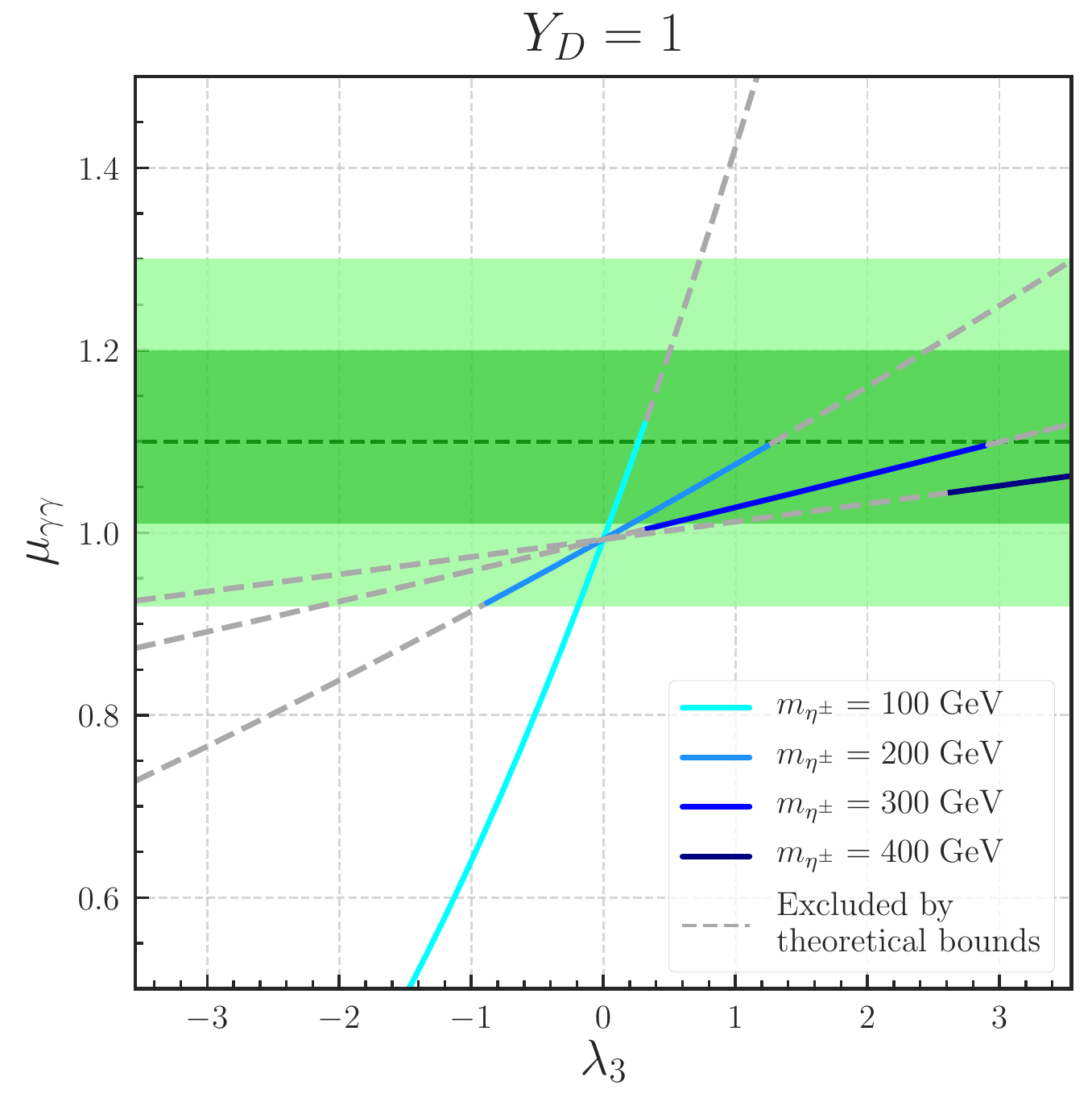}
    \hspace{5mm}
    \includegraphics[width=0.45\textwidth]{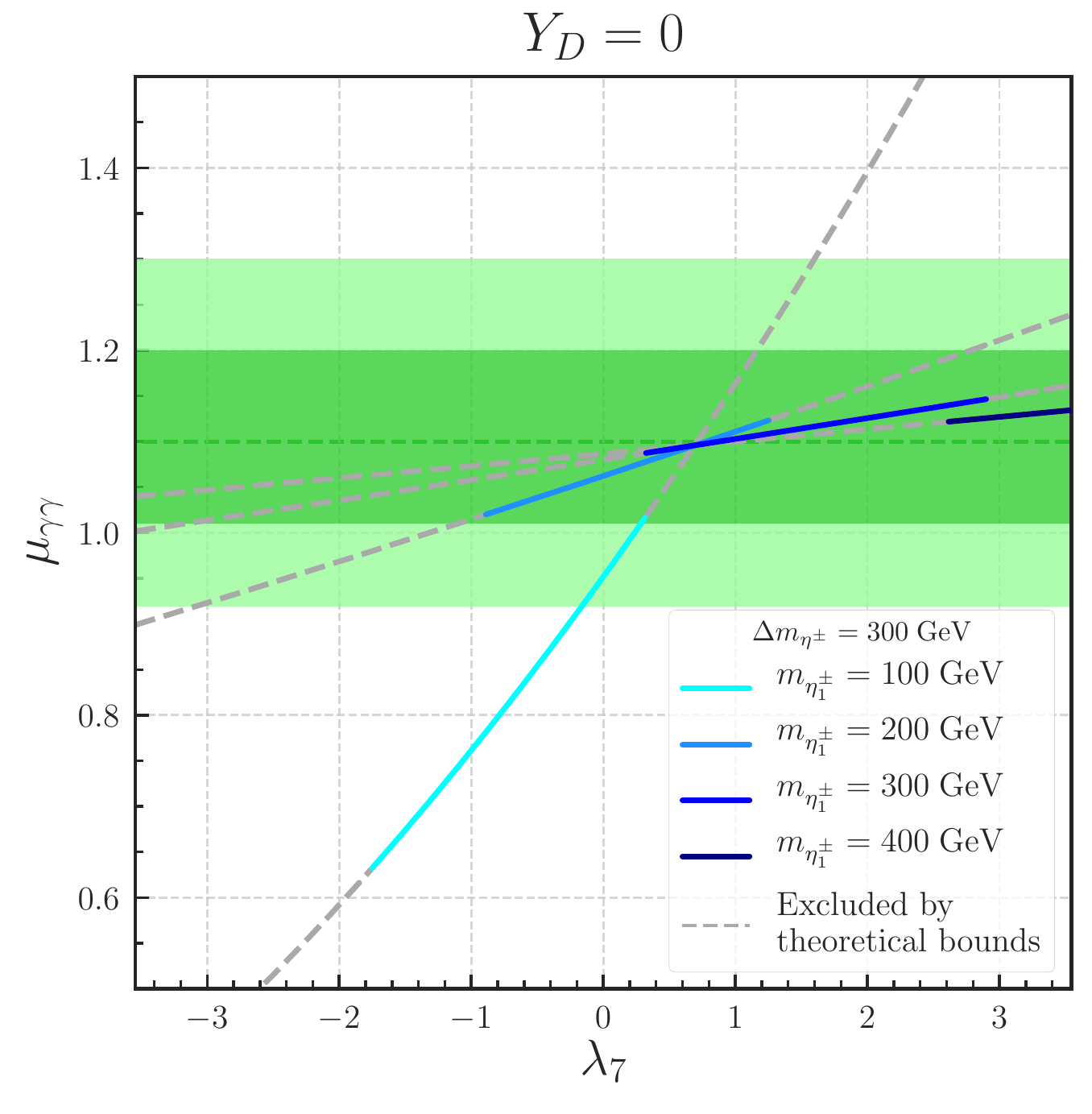}
    \caption{Signal strength $\mu_{\gamma\gamma}$ in the model with $Y_D =1$ (left) and $Y_D=0$ (right). 
The dark (light) green band shows the current global average of $\mu_{\gamma\gamma}^{\rm Exp}$ with $1\sigma$ ($2\sigma$) uncertainty. 
For $Y_D = 0$, we take $(m_{\eta_H}, m_{\eta_A}, \theta) = (80~\mathrm{GeV}, 200~\mathrm{GeV}, \pi/4)$, and the mass splitting between the two charged scalar bosons is fixed to be $300~\mathrm{GeV}$.}
\label{fig:h2aa}
\end{figure}

In Fig.~\ref{fig:h2aa}, we show the signal strength $\mu_{\gamma\gamma}$ as a function of $\lambda_3$ ($\lambda_7$) for different charged scalar masses in the scenario of $Y_D = 1$  $(Y_D = 0)$ in the left (right) panel. For $Y_D = 1$, the scalar boson loops can interfere constructively with the dominant weak gauge boson loops for a negative value of $\lambda_{h\eta^+\eta^-}$ (corresponding to a positive $\lambda_3$). A similar effect is also seen for positive $\lambda_7$ in the right plot with $Y_D = 0$. 
In these plots, the dashed part of each curve is excluded by the perturbative unitarity or vacuum stability bounds according to Eqs.~(\ref{eq:per_bound}) to (\ref{eq:stability2}). 
For $Y_D = 1$, the lower bounds on $\lambda_3$ are determined by the vacuum stability constraints, while 
the upper bounds indirectly come from the vacuum stability constraints on $\lambda_7$ assuming $\lambda_{h\eta_1^0\eta_1^0}/v = 10^{-3}$, as suggested in Sec.~\ref{subsec:DM}. We note that the quartic couplings $\lambda_{2,6,8}$ for the inert scalar fields are scanned for any given $\lambda_3$ such that the allowed range of $\lambda_3$ is maximized. For $Y_D = 0$, the bounds on $\lambda_7$ are derived in a similar way. In this scenario, the lower bounds on $\lambda_7$ arise from the bounded-from-below conditions in Eqs.~(\ref{eq:stability1}) and (\ref{eq:stability2}), while the upper bounds are determined by the requirement $\mu^2_{S} > 0$ [see Eq.~\eqref{eq:potential_param_YD=0}]. 
From Fig.~\ref{fig:h2aa}, it is clear that both scenarios of our model are able to accommodate the current experimental constraints from the $h\rightarrow\gamma\gamma$ decay within a reasonably large range of parameter space without violating the perturbative unitarity and vacuum stability constraints.

\begin{figure}
    \centering
    \includegraphics[width=0.45\textwidth]{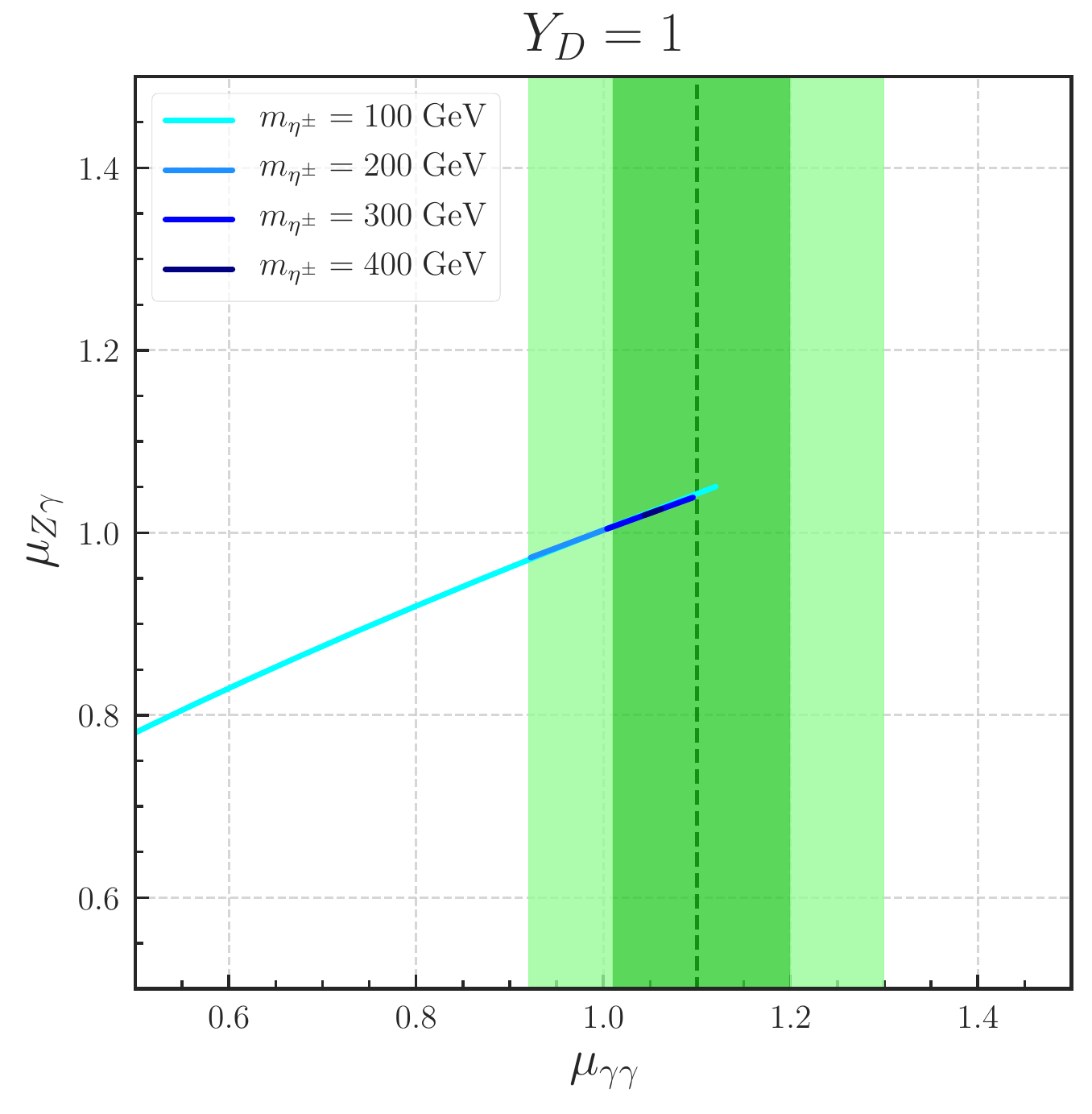}
    \hspace{5mm}
    \includegraphics[width=0.45\textwidth]{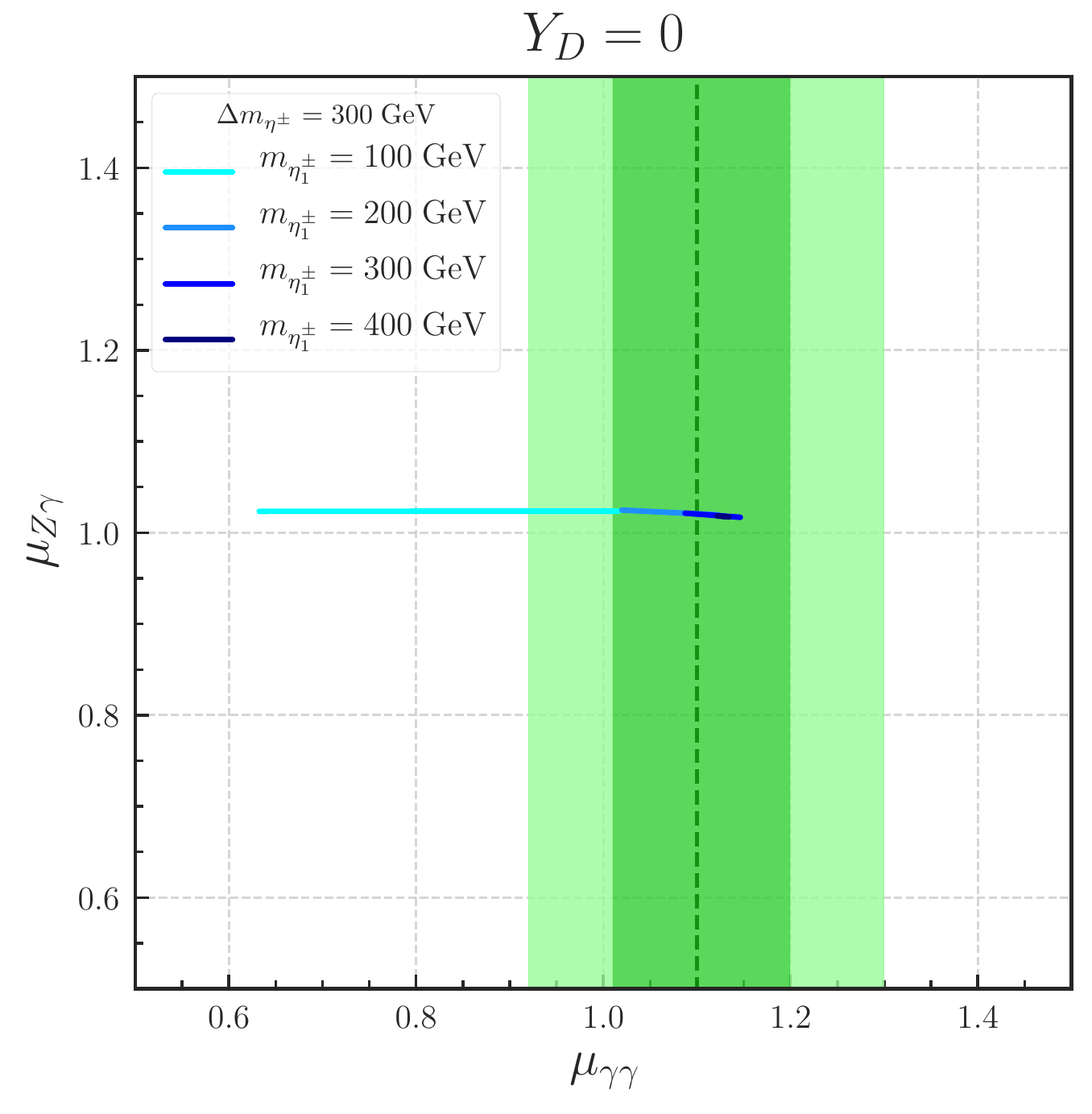}
    \caption{Correlation between $\mu_{Z\gamma}$ and $\mu_{\gamma\gamma}$ in the model with $Y_D =1$ (left) and $Y_D=0$ (right) under the constraints of perturbative unitarity and vacuum stability. 
The dark (light) green band shows the current global average of $\mu_{\gamma\gamma}$ with $1\sigma$ ($2\sigma$) uncertainty. 
For $Y_D = 0$, we take $(m_{\eta_H}, m_{\eta_A}, \theta) = (80~\mathrm{GeV}, 200~\mathrm{GeV}, \pi/4)$, and the mass splitting between the two charged scalar bosons is fixed to be $300~\mathrm{GeV}$.} 
\label{fig:h2Za}
\end{figure}

As the decay rates of $h \to \gamma\gamma$ and $h \to Z\gamma$ have different dependences on couplings, 
to see the correlation between $\mu_{\gamma\gamma}$ and $\mu_{Z\gamma}$ would be useful in order to extract the structure of the model~\cite{Chiang:2012qz}. 
In Fig.~\ref{fig:h2Za}, we show the correlation between $\mu_{Z\gamma}$ and $\mu_{\gamma\gamma}$ for the scenario of $Y_D = 1$ (left) and $Y_D = 0$ (right). 
We only show the points which are allowed by the perturbative unitarity and vacuum stability bounds. 
For $Y_D = 1$, we see that $\mu_{Z\gamma}$ is strongly correlated with $\mu_{\gamma\gamma}$. 
Within the $2\sigma$ region around the current measurements of $\mu_{\gamma\gamma}^{\rm Exp}$, 
a signal strength for $h\to Z\gamma$ is predicted to be from $0.97$ to $1.05$. 
Such a prediction can be slightly modified by the choice of the mixing angle $\theta$ and the masses of the $\mathbb{Z}_2$-odd scalar bosons. 
For $Y_D = 0$, we observe no or little correlation between $\mu_{Z\gamma}$ and $\mu_{\gamma\gamma}$. 
This is because the contributions from the pure $\eta^\pm_1$ and $\eta^\pm_2$ loops are small in our particular choice of $\theta = \pi/4$ due to smaller 
$\eta^+_1\eta^-_1 Z$ and $\eta^+_2\eta^-_2 Z$ couplings.  On the other hand, the $\eta^\pm_1$ and $\eta^\pm_2$ mixed loop contribution, which appears in the $h \to Z\gamma$ decay but not the $h \to \gamma\gamma$ decay, can be sizable. 
The coupling $\lambda_{h\eta_1^\pm\eta_2^\mp}$ that contributes to this new diagram is given by
\begin{equation}
    \lambda_{h\eta^\pm_1\eta^\mp_2} = v s_\theta c_\theta\left[\lambda_3 + \frac{1}{v^2}\left(m^2_{\eta^\pm_1} + m^2_{\eta^\pm_2} - m^2_{\eta_A} - m^2_{\eta_H}\right) - \lambda_7\right]. \label{eq:12}
\end{equation}
With this additional mixed loop contribution, the model with $Y_D = 0$ can predict $\mu_{Z\gamma}\neq 1$ even when $\mu_{\gamma\gamma} = 1$. 
We note that our prediction on $\mu_{Z\gamma}$ is sensitive to the choice of $\theta$, because of the $Z\eta_i^\pm \eta_j^\mp$ couplings. 
By scanning the mixing angle $\theta$ while imposing both theoretical and experimental constraints, 
we find that the model with $Y_D = 0$ would predict an $h \to Z\gamma$ signal strength that is at most $+10\%$ larger than the SM value.

\begin{figure}
    \centering
    \includegraphics[width=0.7\textwidth]{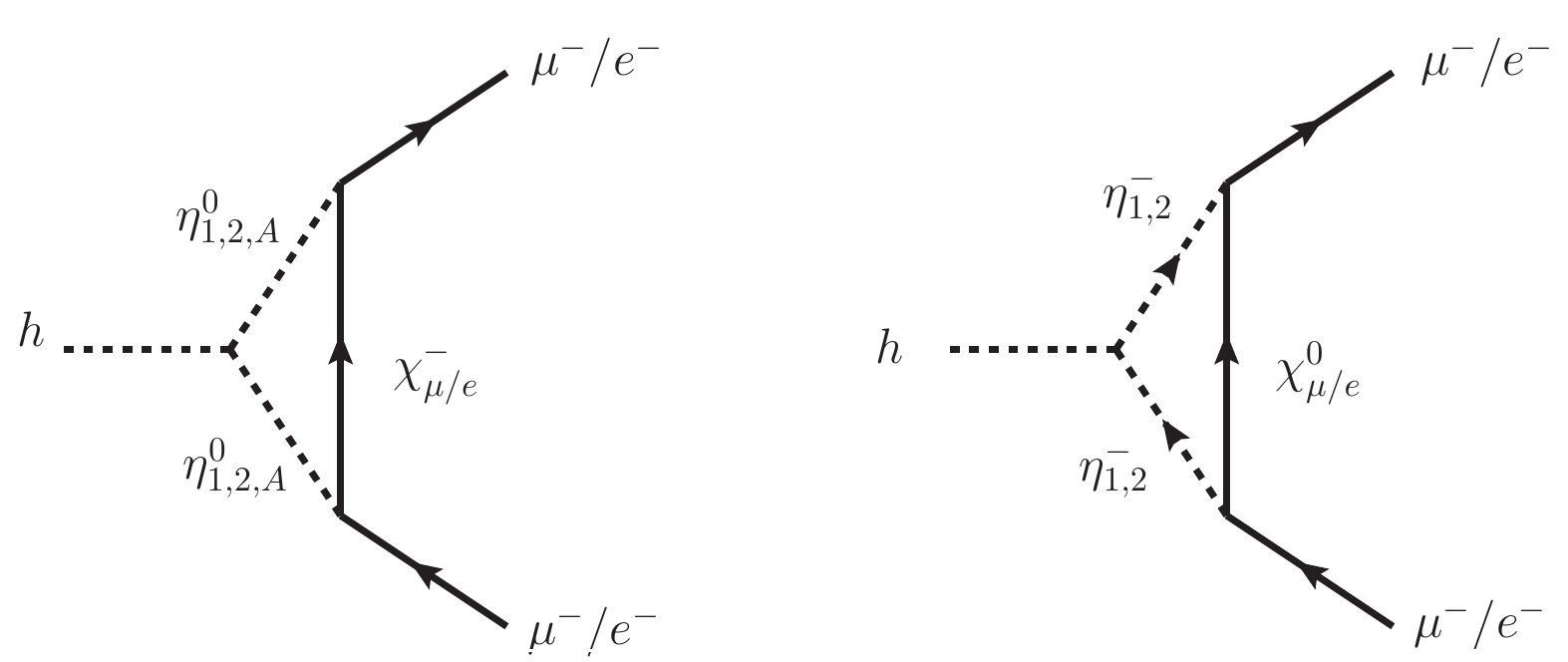}
    \caption{New physics contributions to the $h\ell^+\ell^-$ ($\ell = e,\mu$) vertex in the model with $Y_D = 1$ (left) and $Y_D=0$ (right). } 
\label{fig:hellell}
\end{figure}

{Finally, we discuss the deviations in the $h\mu^+\mu^-$ and $he^+e^-$ couplings from their SM values due to the one-loop corrections from the $\mathbb{Z}_2$-odd particles. 
The new physics contribution to these vertices is shown in Fig.~\ref{fig:hellell}, and the corresponding Yukawa coupling correction is calculated as 
\begin{align}
\Delta y_\ell =& 
\frac{M_{\chi_\ell}}{8\pi^2}\Bigg[\sum_{k=1,2}\lambda_{h\eta_k^0\eta_k^0}g_L^{k,\ell}g_R^{k,\ell}C_0(0,0,m_h^2;m_{\eta_k},M_{\chi_\ell},m_{\eta_k}) \notag\\
&
+\frac{\lambda_{h\eta_1^0\eta_2^0}}{2}(g_L^{1,\ell}g_R^{2,\ell} + g_R^{1,\ell}g_L^{2,\ell})C_0(0,0,m_h^2;m_{\eta_1},M_{\chi_\ell},m_{\eta_2})\Bigg] \quad (\text{for }Y_D = 1), \\
\Delta y_\ell =&
\frac{M_{\chi_\ell}}{8\pi^2}\Bigg[\sum_{k=1,2}\lambda_{h\eta_k^+\eta_k^-}g_L^{k,\ell}g_R^{k,\ell}C_0(0,0,m_h^2;m_{\eta_k^\pm},M_{\chi_\ell},m_{\eta_k^\pm}) \notag\\
&
+\frac{\lambda_{h\eta_1^\pm\eta_2^\mp}}{2}(g_L^{1,\ell}g_R^{2,\ell} + g_R^{1,\ell}g_L^{2,\ell})C_0(0,0,m_h^2;m_{\eta_1^\pm},M_{\chi_\ell},m_{\eta_2^\pm})
\Bigg] \quad (\text{for }Y_D = 0), 
\end{align}
where $C_0$ is the Passarino-Veltman's scalar three-point function~\cite{Passarino:1978jh} (see ref.~\cite{Kanemura:2015mxa} for the definition of the $C_0$ function), and we have neglected the lepton mass in the final state. 
We note that there are also contributions from counter terms, but they are proportional to the muon or electron mass and can be safely neglected.  
For the choice of $\theta = \pi/4$, $f_R^\ell = \sigma_\ell f_L^\ell$ ($\sigma_\ell = +1$ for $\ell = e$ and $\sigma_\ell = -1$ for $\ell =\mu$) 
and $|f_L^\ell| = |f_R^\ell| (= f^\ell)$, the above expressions are simplified to
\begin{align}
\Delta y_\ell & \simeq
\frac{(f^\ell)^2}{16\sqrt{2}\pi^2}\frac{\lambda_{h\eta_1^0\eta_1^0}+\sigma_\ell\lambda_{h\eta_2^0\eta_2^0}}{M_{\chi_\ell}}\left(1 - \ln \frac{M_{\chi_\ell}^2}{m_{\eta^0}^2}\right) \quad (\text{for }Y_D = 1),\\
\Delta y_\ell & \simeq  \frac{(f^\ell)^2}{16\pi^2}\frac{\lambda_{h\eta_1^+\eta_1^-}+\sigma_\ell\lambda_{h\eta_2^+\eta_2^-}}{M_{\chi_\ell}}\left(1 - \ln \frac{M_{\chi_\ell}^2}{m_{\eta^\pm}^2} \right) \quad (\text{for }Y_D = 0), 
\end{align}
where we assumed $M_{\chi_\ell}\gg m_{\eta^0}, \, m_{\eta^\pm}$ with $m_{\eta^0} \equiv m_{\eta_1} (= m_{\eta_2})$ and $m_{\eta^\pm} \equiv m_{\eta_1^\pm} (= m_{\eta_2^\pm})$.
Interestingly, these expressions are not suppressed by the muon or electron mass because these diagrams are controlled by the new Yukawa interaction and the chirality flip happens via the mass of the intermediate vector-like lepton, instead of picking up the external light lepton mass. 
Thus, in spite of being a one-loop process, these contributions can be comparable or even larger than the tree-level one. 
Taking $M_{\chi_\mu} = M_{\chi_e}=1$~TeV, $m_{\eta_1}=80$~GeV, $m_{\eta_2}=380$~GeV, $f^\mu = 0.3$, $f^e = 0.1$, $\lambda_{h\eta_1^0\eta_1^0} = 2.6 \times 10^{-3} v$ and $\lambda_{h\eta_2^0\eta_2^0} = -1.095 v$ for the $Y_D = 1$ scenario as an example, we obtain 
$\Delta y_\mu \simeq 1.63 \times 10^{-4}$ and $\Delta y_e= -1.78 \times 10^{-5}$. 
These correspond to about $+38\%$ and $-858\%$ corrections with respect to the tree-level predictions for the $h\mu^+\mu^-$ and $he^+e^-$ couplings, respectively, which 
can also be considered as the deviation in these couplings from the SM predictions\footnote{It may still be challenging to measure the $he^+e^-$ coupling even if we have such a huge correction because of the tiny electron Yukawa coupling at tree level.}. 
Such a large deviation in the $h\mu^+\mu^-$ coupling can possibly be detected in future collider experiments. For example, at the High-Luminosity LHC (HL-LHC) with the integrated luminosity of 3~ab$^{-1}$ the expected accuracy for measuring the $h \mu^+\mu^-$ coupling is about 14\%~\cite{Fujii:2017vwa}. 
The accuracy can be improved by about 5\%~\cite{Fujii:2017vwa} through the combination of experiments at the HL-LHC and at the 250-GeV International Linear Collider (ILC) with the integrated luminosity of 2~ab$^{-1}$. 
Therefore, our model can be tested by precision measurement of the muon Yukawa coupling with the Higgs boson.
}

\section{Conclusions \label{sec:conclusion}}

To explain the muon and electron $g-2$ anomalies and the dark matter data, we have proposed a new model whose symmetry is enlarged to have a global $U(1)_\ell$ and a discrete $\mathbb{Z}_2$ symmetries and whose particle content is extended with two vector-like leptons and the inert scalar singlet and doublet fields.  Depending upon the hypercharge assignment of the new fields, there are two different scenarios.  Thanks to the new symmetries, we can safely avoid the lepton flavor-violating decays of charged leptons, while obtaining new contributions to the muon and electron $g-2$ with the desired signs and magnitudes for the data.  In addition, the symmetries guarantee the stability of the DM candidate, which is the lightest neutral $\mathbb{Z}_2$-odd particle.

We have found that there are regions in the parameter space that can simultaneously accommodate both $g-2$ anomalies and the DM relic density under the constraints from the LHC direct searches for vector-like leptons and DM direct detection experiments.  In the successful parameter regions, the masses of the vector-like leptons can be about 300~GeV with the magnitude of new muon and electron Yukawa couplings being about $0.1$ and $0.03$, respectively.  Larger vector-like lepton masses generally go with larger values of new Yukawa couplings, while too large values of the Yukawa couplings cause too large annihilation cross section of DM to explain the current observed relic density.  We have shown that typically the magnitude of the new Yukawa couplings should be smaller than about $0.4$. { We have also} discussed the modifications to the Higgs diphoton and Higgs to $Z\gamma$ decays, which are mediated by the inert charged scalar boson loops. We have seen that the predictions of the $h \to \gamma\gamma$ signal strength in our model are mostly consistent with the current measurements at the LHC. Depending on the choice of parameters, our model would further predict an $h \to Z\gamma$ signal strength that is at most $+10\%$ larger than the SM value. 
{Finally, we have discussed the modification to the Higgs couplings with muons and electrons in the model. 
In the viable parameter space, the muon Yukawa coupling can be modified up to about 38\% due to the inert particles running in the loops that are not suppressed by the muon or electron mass.  Such a large deviation can be probed at the HL-LHC and/or the ILC.}

\section*{Acknowledgments}

KFC and CWC were supported in part by the Ministry of Science and Technology (MOST) of Taiwan under Grant No.~MOST-108-2112-M-002-005-MY3. KFC also acknowledges support from the MOST grant MOST-108-2112-M-001-011 and an Academia Sinica grant AS-CDA-106-M01. KY was supported in part by the Grant-in-Aid for Early-Career Scientists, No.~19K14714.

\vspace*{4mm}

%
%

\bibliography{references}

\end{document}